
\input harvmac.tex


\input epsf.tex
\def\figin{\epsfcheck\figin}\def\figins{\epsfcheck\figins}
\def\epsfcheck{\ifx\epsfbox\UnDeFiSIeD
\message{(NO epsf.tex, FIGURES WILL BE IGNORED)}
\gdef\figin##1{\vskip2in}\gdef\figins##1{\hskip.5in}
\else\message{(FIGURES WILL BE INCLUDED)}%
\gdef\figin##1{##1}\gdef\figins##1{##1}\fi}
\def\DefWarn#1{}
\def\figinsert{\goodbreak\midinsert}
\def\ifig#1#2#3{\DefWarn#1\xdef#1{fig.~\the\figno}
\writedef{#1\leftbracket fig.\noexpand~\the\figno}%
\figinsert\figin{\centerline{#3}}\medskip\centerline{\vbox{\baselineskip12pt
\advance\hsize by -1truein\noindent\footnotefont{\bf Fig.~\the\figno:} #2}}
\bigskip\endinsert\global\advance\figno by1}

\lref\rango{S.~Corley, A.~Jevicki and S.~Ramgoolam,
arXiv:hep-th/0111222.
}

\lref\douglasreview{
M.~R.~Douglas and S.~Randjbar-Daemi,
arXiv:hep-th/9902022.
}

\lref\aharonyetal{
O.~Aharony, S.~S.~Gubser, J.~M.~Maldacena, H.~Ooguri and Y.~Oz,
Phys.\ Rept.\  {\bf 323}, 183 (2000)
[arXiv:hep-th/9905111].
}

\lref\divecchia{
P.~Di Vecchia,
arXiv:hep-th/9908148.
}

\lref\coleman{ S. Coleman, ``Aspects of symmetry'', Cambridge
Univ. Press, (1985). }

\lref\metsaev{
R.~R.~Metsaev,
       Nucl. Phys. B {\bf 625}, 70 (2002), arXiv:hep-th/0112044.
}

\lref\giant{
J.~McGreevy, L.~Susskind and N.~Toumbas,
JHEP {\bf 0006}, 008 (2000)
[arXiv:hep-th/0003075].
}

\lref\thooft{
G.~'t Hooft,
Nucl.\ Phys.\ B {\bf 72}, 461 (1974).
}

\lref\polyakov{
A.~M.~Polyakov,
Nucl.\ Phys.\ Proc.\ Suppl.\  {\bf 68}, 1 (1998)
[arXiv:hep-th/9711002].
}

\lref\review{
O.~Aharony, S.~S.~Gubser, J.~Maldacena, H.~Ooguri and Y.~Oz,
Phys.\ Rept.\  {\bf 323}, 183 (2000)
[arXiv:hep-th/9905111].
}

\lref\gkp{
S.~S.~Gubser, I.~R.~Klebanov and A.~M.~Polyakov,
Phys.\ Lett.\ B {\bf 428}, 105 (1998)
[arXiv:hep-th/9802109].
}

\lref\jm{
J.~Maldacena,
Adv.\ Theor.\ Math.\ Phys.\  {\bf 2}, 231 (1998)
[Int.\ J.\ Theor.\ Phys.\  {\bf 38}, 1113 (1998)]
[arXiv:hep-th/9711200].
}

\lref\igorreview{
I.~R.~Klebanov,
arXiv:hep-th/9901018.
}

\lref\wittenhol{
E.~Witten,
Adv.\ Theor.\ Math.\ Phys.\  {\bf 2}, 253 (1998)
[arXiv:hep-th/9802150].
}

\lref\polyakov{
A.~M.~Polyakov,
arXiv:hep-th/0110196.
}

\lref\vijay{
V.~Balasubramanian, M.~Berkooz, A.~Naqvi and M.~J.~Strassler,
arXiv:hep-th/0107119.
}

\lref\longstrings{
N.~Seiberg and E.~Witten,
JHEP {\bf 9904}, 017 (1999)
[arXiv:hep-th/9903224].
 ;J.~Maldacena and H.~Ooguri,
J.\ Math.\ Phys.\  {\bf 42}, 2929 (2001)
[arXiv:hep-th/0001053].
}

\lref\bfss{T.~Banks, W.~Fischler, S.~H.~Shenker and L.~Susskind,
Phys.\ Rev.\ D {\bf 55}, 5112 (1997)
[arXiv:hep-th/9610043].
}
\lref\susskind{L.~Susskind,
arXiv:hep-th/9704080.
}
\lref\sen{A.~Sen,
Adv.\ Theor.\ Math.\ Phys.\  {\bf 2}, 51 (1998)
[arXiv:hep-th/9709220].
}
\lref\seiberg{N.~Seiberg,
Phys.\ Rev.\ Lett.\  {\bf 79}, 3577 (1997)
[arXiv:hep-th/9710009].
}

\lref\bsei{
T.~Banks and N.~Seiberg,
Nucl.\ Phys.\ B {\bf 497}, 41 (1997)
[arXiv:hep-th/9702187].
}

\lref\matrixmodels{
E.~Brezin and V.~A.~Kazakov,
Phys.\ Lett.\ B {\bf 236}, 144 (1990).
M.~R.~Douglas and S.~H.~Shenker,
Nucl.\ Phys.\ B {\bf 335}, 635 (1990).
D.~J.~Gross and A.~A.~Migdal,
Phys.\ Rev.\ Lett.\  {\bf 64}, 127 (1990).
D.~J.~Gross and N.~Miljkovic,
Phys.\ Lett.\ B {\bf 238}, 217 (1990).
E.~Brezin, V.~A.~Kazakov and A.~B.~Zamolodchikov,
Nucl.\ Phys.\ B {\bf 338}, 673 (1990).
}

\lref\ppwavereview{
J.~C.~Plefka,
arXiv:hep-th/0307101.
C.~Kristjansen,
arXiv:hep-th/0307204.
D.~Sadri and M.~M.~Sheikh-Jabbari,
arXiv:hep-th/0310119.
}

\lref\fp{
J.~Figueroa-O'Farrill and G.~Papadopoulos,
supersymmetric solution of  M-theory,''
JHEP {\bf 0108}, 036 (2001)
[arXiv:hep-th/0105308].
}
\lref\figueroaiib{
M.~Blau, J.~Figueroa-O'Farrill, C.~Hull and G.~Papadopoulos,
JHEP {\bf 0201}, 047 (2001)
[arXiv:hep-th/0110242].
}

\lref\figueroarecent{
M.~Blau, J.~Figueroa-O'Farrill, C.~Hull and G.~Papadopoulos,
arXiv:hep-th/0201081.
}

\lref\penrose{R. Penrose, 
Differential geometry and relativity, Reidel, Dordrecht, 1976, pp.
271-275.
}

\lref\kg{J.~Kowalski-Glikman,
Phys.\ Lett.\ B {\bf 134}, 194 (1984).
}

\lref\planewaves{See for example:
D.~Amati and C.~Klimcik,
Phys.\ Lett.\ B {\bf 210}, 92 (1988).
G.~T.~Horowitz and A.~R.~Steif,
Phys.\ Rev.\ Lett.\  {\bf 64}, 260 (1990).
H.~J.~de Vega and N.~Sanchez,
Phys.\ Lett.\ B {\bf 244}, 215 (1990);
In String Theory And String Propagation
Through Gravitational Shock Waves,''
Phys.\ Rev.\ Lett.\  {\bf 65}, 1517 (1990).
O.~Jofre and C.~Nunez,
Phys.\ Rev.\ D {\bf 50}, 5232 (1994)
[arXiv:hep-th/9311187].
}

\lref\polchinski{
J.~Polchinski,
``String Theory''
{\it  Cambridge, UK: Univ. Pr. (1998)}.
}

\lref\ghmnt{ M.~T.~Grisaru, P.~Howe, L.~Mezincescu, B.~E.~W.~Nilsson,
P.~K.~Townsend, 
Phys.\ Lett.\ B {\bf 162}, 116 (1985)
}

\lref\wpps{ B.~de~Wit, K.~Peeters, J.~Plefka, A.~Sevrin,
Phys.\ Lett.\ B {\bf 443}, 153 (1998)
[arXiv:hep-th/9808052]
}

\lref\krr{ R.~Kallosh, J.~Rahmfeld, A.~Rajaraman,
JHEP {\bf 9809}, 002 (1998)
[arXiv:hep-th/9805217]
}

\lref\myers{ R. Myers, 
JHEP {\bf 12}(1999) 022, hep-th/9910053
}

\lref\kt{ R.~Kallosh, A.~A.~Tseytlin,
JHEP {\bf 9810}, 016 (1998)
[arXiv:hep-th/9808088]
}

\lref\mt{ R.~R.~Metsaev, A.~A.~Tseytlin,
Phys.\ Rev.\ D {\bf 63}, 046002 (2001)
[arXiv:hep-th/0007036]
}

\lref\mtt{ R.~R.~Metsaev, C.~B.~Thorn, A.~A.~Tseytlin,
Nucl.\ Phys.\ B {\bf 596}, 151 (2001)
[arXiv:hep-th/0009171]
}

\lref\dkss{ S.~Deger, A.~Kaya, E.~Sezgin, P.~Sundell,
Nucl.\ Phys.\ B {\bf 536},
110 (1998)
[arXiv:hep-th/9804166]
}

\lref\meessen{ P.~Meessen,
arXiv:hep-th/0111031.
}

\lref\ps{J.~Polchinski and M.~J.~Strassler,
arXiv:hep-th/0003136.
}
\lref\motl{
L.~Motl,
arXiv:hep-th/9701025.
}
\lref\dvv{
R.~Dijkgraaf, E.~Verlinde and H.~Verlinde,
Nucl.\ Phys.\ B {\bf 500}, 43 (1997)
[arXiv:hep-th/9703030].
}

\lref\sfetsos{K.~Sfetsos,
Phys.\ Lett.\ B {\bf 324}, 335 (1994)
[arXiv:hep-th/9311010].
K.~Sfetsos and A.~A.~Tseytlin,
Nucl.\ Phys.\ B {\bf 427}, 245 (1994)
[arXiv:hep-th/9404063].
}

\lref\guven{R.~Gueven,
Phys.\ Lett.\ B {\bf 191}, 275 (1987).
 R. Guven,  ``Plane Wave Limits and T-duality'', Phys. Lett. B482(2000)
      255.}

\lref\bmn{
D.~Berenstein, J.~Maldacena and H.~Nastase,
arXiv:hep-th/0202021.
}

\lref\gsw{
M.~B.~Green, J.~H.~Schwarz and E.~Witten,
``Superstring Theory ''
{\it  Cambridge, Uk: Univ. Pr. ( 1987) 469 P. ( Cambridge Monographs On Mathematical Physics)}.
}

\lref\ms{
R.~R.~Metsaev and A.~A.~Tseytlin,
arXiv:hep-th/0202109.
}

\lref\figueroanew{
M.~Blau, J.~Figueroa-O'Farrill and G.~Papadopoulos,
arXiv:hep-th/0202111.
}

\lref\recent{
E.~Floratos and A.~Kehagias,
arXiv:hep-th/0203134.
U.~Gursoy, C.~Nunez and M.~Schvellinger,
arXiv:hep-th/0203124.
M.~Billo' and I.~Pesando,
arXiv:hep-th/0203028.
M.~Alishahiha and M.~M.~Sheikh-Jabbari,
arXiv:hep-th/0203018.
L.~A.~Zayas and J.~Sonnenschein,
arXiv:hep-th/0202186.
J.~G.~Russo and A.~A.~Tseytlin,
arXiv:hep-th/0202179.
J.~Gomis and H.~Ooguri,
arXiv:hep-th/0202157.
N.~Itzhaki, I.~R.~Klebanov and S.~Mukhi,
arXiv:hep-th/0202153.
J.~Michelson,
arXiv:hep-th/0203140.
}




\Title{\vbox{\baselineskip12pt \hbox{hep-th/0309246}  }}
{\vbox{
\centerline{TASI 2003 lectures on AdS/CFT}
}}
\smallskip
\centerline{ Juan Maldacena}
\medskip

\smallskip

\centerline{\it  Institute for Advanced Study} \centerline{\it
Princeton, New Jersey 08540, USA}
\smallskip

\bigskip
\noindent

We give a short introduction to AdS/CFT and its plane wave limit.

\Date{September 2003}

\vfil\eject


\newsec{Introduction}

In these lecture notes we provide a short introduction to
the ideas related to the correspondence between gauge theories
and gravity theories. For other reviews of the subject, includding
a more complete list of references,
see
\refs{\igorreview,\douglasreview,\divecchia,\aharonyetal,\ppwavereview}.

We start by discussing the simplifications
that ocurr in the large $N$ limit of field theories. We discuss
first the large $N$ limit of vector theories, then the large
$N$ limit of theories where the fundamental fields are $N\times N$
matrices and we show that these theories are expected to
be described in terms of strings
\ref\tHooft{
G.~'t Hooft,
Nucl.\ Phys.\ B {\bf 72}, 461 (1974).
}.
If we start with a four dimensional gauge theory, we might
naively expect to find  a  strings moving in four dimensions.
 But strings are not consistent in four flat dimensions.
As we try to proceed,
we are forced to introduce at least one more dimension
\ref\PolyakovTJ{
A.~M.~Polyakov,
Nucl.\ Phys.\ Proc.\ Suppl.\  {\bf 68}, 1 (1998)
[arXiv:hep-th/9711002].
}.
If the gauge theory is conformal, then the original flat dimensions,
together with this new extra dimension are constrained by the
symmetries to form an Anti-de-Sitter spacetime. We will describe
some basic properties of Anti-de-Sitter spacetimes. Then
we present the simplest example of the relationship between
a four dimensional field theory and a gravity theory. Namely,
we discuss the relationship between Yang Mills theory with
four
supersymmetries to type IIB superstring theory on $AdS_5 \times
S^5$ \jm .
We later give the general prescription linking computations of
correlation functions in the gauge theory to the computations
of amplitudes in the gravity theory \gkp \wittenhol .
This is a general prescription that should hold for any field
theory that has a gravity dual.

Finally we discuss a particular limit of the relationship
between ${\cal N} =4$ Yang Mills and $AdS_5 \times S^5$ where
we consider particles with large angular momentum on the
sphere \bmn . In this limit the relevant region of the geometry
looks like a plane wave where we can quantize strings exactly.
Through a simple gauge theory computation one can reproduce
the string spectrum.

\newsec{ Large $N$ }

There are theories that contain a large number
of fields related
by a symmetry such as $SO(N)$ or $U(N)$. These
theories simplify when $N$ is taken to infinity.
For a more extended discussion of this
subject see \coleman .

\subsec{ Large $N$ for vector theories}

Consider a theory with $N$ fields $\eta^i$, where $i=1, \cdots N$
with O(N) symmetry. For example, \eqn\on{ S= { 1 \over 2 g_0^2}
\int d^2 \sigma (\partial \vec n)^2 ~,~~~~~~~ \vec n^2 =1 } First
note that the effective coupling of the theory is $g_0^2 N$. The
theory simplifies in the limit where we keep $g_0^2 N$ fixed and
we take $N \to \infty$. In this limit only a subset of Feynman
diagrams survives.
A very convenient way to proceed is to introduce a
Lagrange multiplier, $\lambda$,
 that enforces the constraint in \on\ and
integrate out the fields $\vec n$.
\eqn\act{\eqalign{
S =& {1 \over 2  g_0^2} \int d^2 \sigma [(\partial \vec n)^2 + \lambda (
\vec n^2 -1)]
\cr
S = & {N\over 2 } \left[ \log\det(-\partial^2 + \lambda )
 - { 1 \over g_0^2 N } \int \lambda d^2 \sigma \right]
}}
We get a classical theory for $\lambda$ in this large $N$ limit,
so we set
 ${\partial S \over \partial \lambda } =0$. We get
 \eqn\getsol{\eqalign{
 1 &=  N g_0^2 \int { d^2p \over (2 \pi)^2} { 1 \over p^2 + \lambda^2 }
 = { N g_0^2 \over 4 \pi} \log \Lambda^2/\lambda
\cr
\lambda = & \Lambda^2 e^{ - { 4 \pi \over N g_0^2 }} = \mu^2 e^{ - {4 \pi
\over Ng^2} }
}}
where $g_0$ is the bare coupling, by absorbing the $\Lambda$
dependence in $g_0$  we define
the renormalized coupling $g$. Notice that the cutoff dependence
of $g_0$ is that of an asymptotically free theory.
By looking again at \act\ we find that the expectation
value for $\lambda$ in \getsol\ gives a mass to the $\vec n $ fields.
Moreover, the model has an unbroken $O(N)$ symmetry. The fact
the O(N) symmetry is restored is consistent with the fact
that in two dimensions we cannot break continuous symmetries.
Note that the $g^2 N$ dependence of $\lambda$ in \getsol\ implies
that the mass for $ \vec n $ is non-perturbative in $g^2 N$.
Notice that, even though the dependence of the mass in $g^2N$ looks
non-perturbative, we have obtained this result by summing Feynman
diagrams, in particular we obtained it through a one loop diagram
contribution to the effective action and then balancing this term
against a tree level term. Large $N$ was crucial to ensure that
no further diagrams contribute.

This theory is similar to $QCD_4$ since it is asymptotically free and
has a mass gap. It has a large $N$ expansion and the large $N$ expansion
contains the fact that the theory has a mass gap. This mass gap
is non perturbative in $g^2 N$.

\subsec{ Matrix theories}

Consider theories where the basic field is a hermitian
matrix $M$. This arises, for example, in a $U(N)$
gauge theory, or a $U(N)$ gauge theory with matter fields in the
adjoint representation.
The Lagrangian has a schematic form
\eqn\lagram{
L = { 1 \over g^2 } Tr[ ( \partial M)^2 + M^2 + M^3 + \cdots ] =
{ 1 \over g^2 } Tr[ ( \partial M)^2 + V(M) ]
}
The action is $U(N)$ invariant $ M \to U M U^\dagger $.
It is convenient to introduce a double line notation
to keep track of the matrix indices
\ifig\figtwo{ Propagator}
{\epsfxsize .7 in\epsfbox{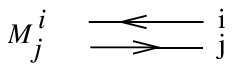}}
\ifig\figthree{Vertices}
{\epsfxsize 1.3 in\epsfbox{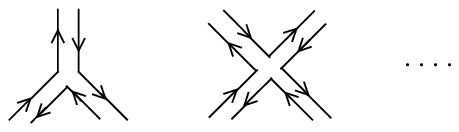}}
Each propagator,  \figtwo, contributes a factor $g^2$
in the Feynman diagrams.
Each vertex,  \figthree, contributes a factor of $1/g^2$.
Finally, each closed line will contain a sum over the gauge
index and will contribute a factor of $N$, see Fig. 3.
\ifig\figfour{ Closed line contributes a factor of $N$.}
{\epsfxsize 1.5 in\epsfbox{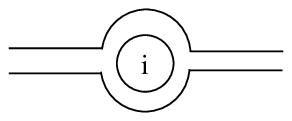}}
Each diagram contributes with
\eqn\contrdia{
(g^2)^{\rm  \# Propagators - \# vertices } N^{ \rm \# Closed ~lines}
}
We can draw these diagrams on a two dimensional surface and
 think of it as a
geometric figure. We see that \contrdia\ becomes
\eqn\contr{
N^{ \rm \# Faces - \# Edges + \# vertices} ( g^2 N)^{\rm Power}
= N^{ 2 - 2 h} (g^2 N)^{ \rm Power }
}
where $h$ is the genus of the two dimensional surface. Namely,
a sphere has genus $h=0$, a torus has genus $h=1$, etc.
\ifig\figfive{Planar diagrams }
{\epsfxsize 2 in\epsfbox{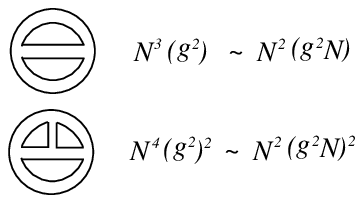}}
\ifig\figsix{A non-planar diagram}
{\epsfxsize 2 in\epsfbox{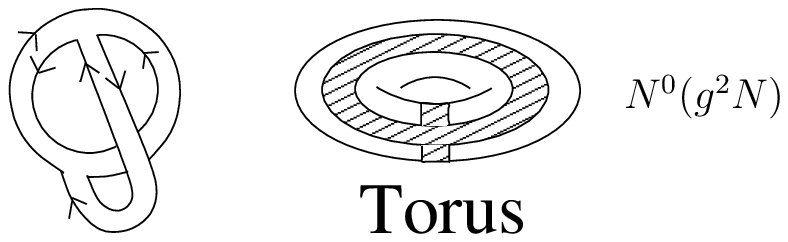}}
A few examples of diagrams that can be drawn on a plane or a sphere
are shown in  \figfive , and example of a diagram that
cannot be drawn on a sphere but can be drawn on a torus is shown
in  \figsix .
The sum of all planar diagrams gives
\eqn\sumpla{
N^2 [ c_0  + c_1 (g^2 N) + c_2 (g^2 N)^2 + \cdots ] = N^2 f(g^2 N)
}
where the $c_i$ are numerical coefficients depending on the
detailed evaluation of each Feynman graph. This detailed
evaluation contains the momentum integrals.
The full partition function has the form
\eqn\partfn{
\log Z = \sum_{h=0}^\infty N^{ 2 - 2 h} f_h( g^2 N)
}
The 't Hooft limit is
\eqn\limit{
N \to \infty ~,~~~~~~~ \lambda \equiv g^2 N = {\rm fixed}
}
$\lambda$ is the 't Hooft coupling.
In this limit only the planar diagrams contribute.
As $\lambda$ gets large a large number of diagrams contribute and
 they become dense on the sphere, so we might think that
they describe a discretized
worldsheet of some string theory. This worldsheet theory is defined
to be whatever results from summing the planar diagrams.
This argument is valid for any matrix theory.
The argument does not give us a practical way of finding the
worldsheet theory.
In bosonic Yang Mills theory $g^2$ runs. In fact, the beta function
has a smooth large $N$ limit
\eqn\betafn{
\dot \lambda = \beta (\lambda) + o(1/N^2)
}
So we have $\lambda(E)$. The string description will be appropriate
where $\lambda(E)$ becomes large. If we add matter in the fundamental,
then we get diagrams with boundaries. These give rise to open strings
which are mesons containing a quark and anti-quark.

Some features of QCD with $N=3$ are similar to those of $N=\infty$,
like the fact that mesons contain a quark and anti-quark, and that
they have small interactions. Strings are also suggested by the
existence of
Regge trajectories. Namely that
 particles with highest spin for a given
mass obey $\alpha' m^2 =  J + {\rm const} $.
 Confinement
is also  closely associated to  a string that forms between the
quark and anti-quark. Though we will see later
that the string description does not necessarily imply confinement.

\subsec{Large $N$ correlators}

Consider operators of the form
\eqn\operat{
{\cal O} = N tr[ f(M) ]
}
diagrammatically represented in Fig. 6.
\ifig\figseven{Operator insertion  }
{\epsfxsize 1 in\epsfbox{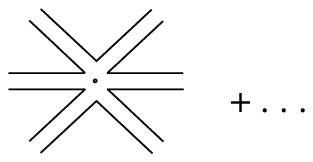}}
If we add them to the action,  they have the same scaling
as  an extra interaction vertex.
In the large $N$ limit their correlation functions factorize
\eqn\factor{
\langle  tr[ f_1(M) ]  tr[ f_2(M)] \rangle  = \langle
 tr[ f_1(M)] \rangle \langle tr[ f_2(M)] \rangle
+ o(1/N^2)
}
Notice that this implies that the leading
contribution is a disconnected
diagram.
All connected correlation functions of
operators normalized as in \operat\
go like $N^2$.
This means that
\eqn\normali{
 \langle {\cal O O} \rangle_c \sim N^2
  ~,~~~~~ \langle {\cal O O O } \rangle_c \sim N^2
  ~,~~~~~~~
{  \langle {\cal O O  O } \rangle_c \over
 \langle {\cal  O O } \rangle_c^{3/2} } \sim { 1 \over N}
 }
 where the subscript indicates the connected part.
In the string description the insertion of these operators
corresponds to the insertion of a vertex operator on the
string worldsheet.

An interesting operator in gauge theories
is the Wilson loop operator
\eqn\loopop{
W({\cal C}) = N Tr[ P e^{ \oint_{\cal C} A } ]
}
For a contour of large area the expectation value of this operator
should go like $e^{- T ({\rm Area}) }$ for a confining theory. $T$ is
then the string tension.

\newsec{ Guessing  the  string theory}

Rather than summing all Feynman diagrams one would like to
guess what the final string theory description is.
Naively, for $d=4$ Yang Mills we expect to  get a bosonic
string theory
that lives in four dimensions.
We know this is not correct.
The bosonic string is not consistent quantum mechanically in $d=4$.
It is consistent in $d=26$ flat dimensions, but this is not the
theory we are interested in.

The reason for this inconsistency is that
the classical Polyakov action
\eqn\polyak{
S \sim  \int \sqrt{g} g^{ab} \partial_a X \partial_b X
}
has a Weyl symmetry $ g_{ab} \to \Omega g_{ab}$ which
is not a symmetry quantum mechanically. In the quantum theory,
under a
change  metric of the form $g_{ab} = e^ {  \phi } \hat g_{ab}
$
the partition function
\eqn\changpart{
e^{ - S_{eff}(g) } = \int DX D(bc) e^{- S[X,g] - S[b,c,g] }
}
changes as
\eqn\liouv{
S_{eff}(g) - S_{eff}(\hat g) = {( 26-D )
\over 48 \pi}  \int { 1 \over 2}  ( \hat \nabla \phi)^2
+ \hat R^{(2)} \phi + \mu^2 e^{ \phi}
}
This action for $\phi$ is called  ``Liouville" action. Even
though the initial classical action for the conformal factor
in the metric was zero, a non-trivial action was generated
quantum mechanically.
Integrating over $\phi$ is like adding a new dimension.

For $D \leq 1$ this is the right answer.
We start with a matrix integral or a matrix quantum
mechanics and we get a string in one or two dimensions.
Actually, it  is necessary to
do a particular scaling limit in the matrix quantum mechanics which
involves $N\to \infty $ and a tuning of
the potential that is analogous
to taking the 't Hooft coupling to a region where there is a large
number of Feynman diagrams that contribute, see \matrixmodels .

For $D=4$ it is not known how to quantize the Liouville action.
Nevertheless the lesson we extract is that we need to include
at least one extra dimension.
So we introduce an extra
dimension and look for the most general string
theory.
If we are interested in four dimensional gauge theories
we look for strings in five dimensions.
We need to specify  the space where string moves.
It should have 4d Poincare symmetry, so the metric has the form
\eqn\spacepoi{
ds^2 = w(z)^2 ( dx^2_{1+3} + d z^2)
}
we have used the reparametrization symmetry to set the coefficient
of $dz^2$ equal to that of $dx^2$.

Now suppose that we were dealing with a scale
invariant field theory.
${\cal N} =4$ Yang Mills is an example.
Then
\eqn\scaling{
x \to \lambda x
}
 should be a symmetry. But in string theory we
have a scale, set by the string tension. So the only way that a
string (with the usual Nambu action\foot{
It is possible to write a string action that is conformal invariant
in four dimensions \ref\PolyakovCS{
A.~M.~Polyakov,
Nucl.\ Phys.\ B {\bf 268}, 406 (1986).
} but it is not know how to quantize it.})
 could be symmetric under
\scaling\  is that this scaling is an isometry of \spacepoi .
This means that $z \to \lambda z$ and that $w = R/z $.
So we are dealing with  5 dimensional Anti-de-Sitter
space
\eqn\adsfive{
ds^2 = R^2 { dx^2 + dz^2 \over z^2 }
}
\ifig\figeight{A sketch of Anti-de-Sitter space. We emphasize
the behavior of the warp factor.   }
{\epsfxsize 3 in\epsfbox{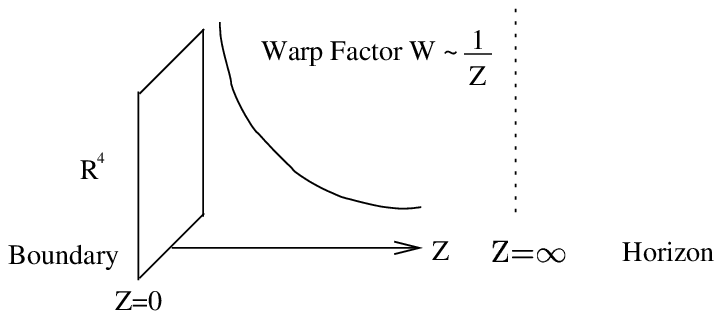}}
This is a spacetime with constant negative curvature and it
is  the most
symmetric spacetime with  negative curvature. The most
symmetric spacetime with positive curvature is de-Sitter.
In Euclidean space the most symmetric positive curvature space is
a sphere and the most symmetric negative curvature one is hyperbolic
space. These are the Euclidean continuation of
de-Sitter and Anti-de-Sitter respectively.

\subsec{Conformal symmetry}

A local field theory that is scale invariant is usually also conformal
invariant.
The change in the action due to a change in the
metric is
\eqn\changeac{
\delta S = \int T^{\mu\nu} \delta g_{\mu \nu}
}
Under a coordinate transformation $x^\mu \to x^\mu + \zeta^\mu(x)$
the action changes by \changeac\
with
\eqn\changemet{
\delta g_{\mu \nu} = \nabla_\mu \zeta_\nu + \nabla_\nu \zeta_\mu
}

If $\zeta^\mu$ generates an isometry then the metric is left
invariant,   so that we
have $\delta g_{\mu\nu} =0$ in \changemet.
The scale transformation \scaling\ gives $\delta g_{\mu \nu} =
2 \delta \lambda
g_{\mu \nu} $. The action would be invariant  if
$T^\mu_\mu =0$. In this case the action is also invariant under
coordinate transformations such that $\delta g_{\mu\nu} = h(x) g_{\mu \nu}
$ in \changemet , and $h$ is any function. Coordinate transformations
of this type are called conformal transformations.
In d=4 they form the group $SO(2,4)$. This group is obtained by
adding to the Poincare group the scale transformation and
the inversion $\vec x \to - \vec x/x^2 $. We see that the inversion
maps the origin to infinity. It turns out that
the conformal group acts more  nicely if
we compactify the space and we
consider $S^3 \times R$ in the Lorentzian
case or $S^4$ in the Euclidean case.

Note also that if the trace of the stress tensor is zero, then the
theory is also Weyl invariant, it is invariant under a rescaling
of the metric $g \to \Omega^2 g$.
In the quantum case this symmetry will have a calculable anomaly and
one can  find the change in physical quantities under such a
rescaling.

\subsec{ Isometries of AdS}

In order to see clearly the AdS isometries
we write $AdS$ as a
hypersurface in $R^{2,4}$
\eqn\adswr{
-X_{-1}^2 - X_0^2 + X_1^2 + X_2^2 + X_3^2 + X_4^2 = - R^2
}
Note that even though the ambient space has 2 time directions the
surface contains only one time direction, the other is orthogonal to the
surface. You should not be confused by these two times, AdS is an ordinary
Lorentzian space with one time!.
We recover \adsfive\ after writing $ X_{-1} + X_4 = R/z$,
$X_\mu = R x_\mu/z$ for $\mu =0, \cdots 3$.
By choosing an appropriate parameterization of \adswr\ we can
also
 write the metric
\eqn\globalme{
ds^2 = R^2 [ - \cosh^2 \rho d\tau^2 +
d \rho^2 + \sinh^2 \rho d \Omega_3^2 ]
}
These are called  ``global" coordinates. They cover the whole
AdS space. In contrast the ``Poincare" coordinates in \adsfive\
only cover a portion. Note that translations in $\tau$ correspond
to rotations of $X_{-1}$ and $X_0$ in \adswr . So in the construction
based on \adswr\ we would get closed timelike curves. Fortunately
we can go to the covering space and consider \globalme\ with
$\tau $ non-compact. When we talk about $AdS$ we are always going to
think about  this covering space.

In the metric \globalme\ we can take out a factor of $\cosh^2 \rho$, and
define a new coordinate $dx = d\rho/\cosh\rho $. We see that the
range of $x$ is finite.
This allows us to understand the Penrose diagram of $AdS$. It is
a solid cylinder whose boundary is $S^3 \times R$ where $R$
corresponds to the time direction. The field theory will be defined
on this boundary. On this boundary the isometries of $AdS$ act like
the conformal group acts in four dimensions.
The proper distance to the boundary along a surface of constant
time is infinite.

\ifig\fignine{(a) Penrose diagram of AdS. (b) Trajectory of
a light ray. (c) Trajectory of massive geodesics.  }
{\epsfxsize 3 in\epsfbox{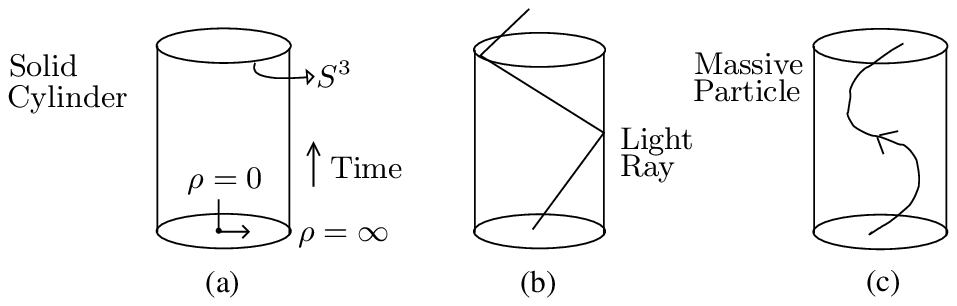}}

\ifig\ftermsfig{ The coordinates of \adsfive\ cover only the
region of global AdS contained between the two shaded hyperplanes.
These hyperplanes correspond to the horizons at $z=\infty$ and
$t = \pm \infty$.    }
{\epsfxsize 2 in\epsfbox{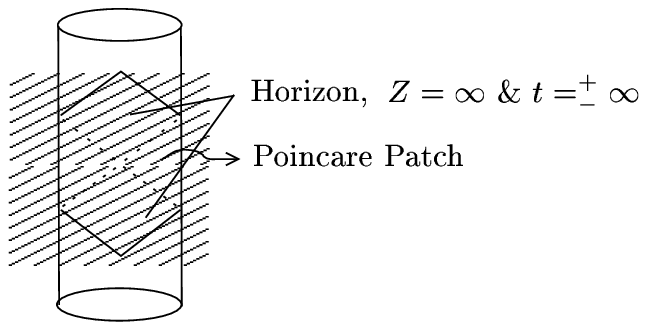}}

Finally note that Weyl transformations in the 4d theory correspond
to picking different functions as conformal factors when we compute
the Penrose diagram so that the boundary will have different metrics
which differ by an
overall function of the coordinates on the boundary.

\subsec{ Mapping of states and operators}

\ifig\figeleven{ We can map states of the field theory on
$S^3 \times R $ to operators on $R^4$.}
{\epsfxsize 3 in\epsfbox{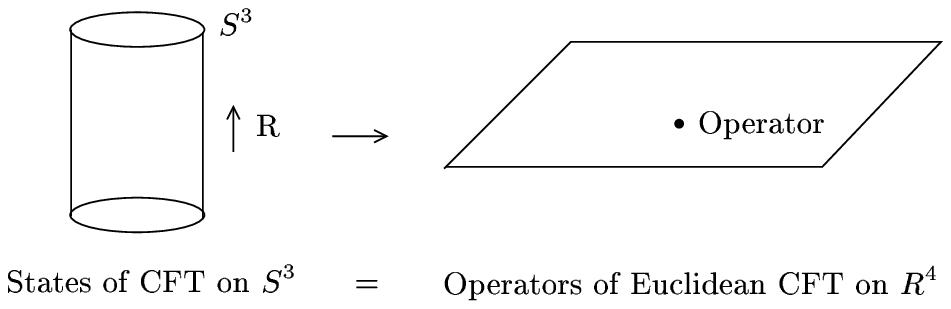}}

In a CFT we have a correspondence between operators on $R^4$ and
states on the cylinder $S^3 \times R$. This can be seen as
follows.
We start with a state on the cylinder,
we go to Euclidean time and then
notice that the cylinder and the plane differ only by a
Weyl transformation
so that the two theories are related.
The vacuum on the cylinder corresponds to the identity on the plane.
The energy of the state in the cylinder corresponds to the conformal
dimension of the operator on the plane, $E_{cyl} = \Delta $.

\subsec{ ${\cal N} =4$  U(N) Yang Mills and strings on
$AdS_5 \times S^5$ }

Consider a theory with four supersymmetries in 4 dimensions,
namely
sixteen real supercharges.
This theory has a unique field content,  there is a unique
supermultiplet. Our only freedom is  the choice of gauge group
and coupling constants.
The field content is as follows. One vector field or gauge boson
$A_\mu$, six scalars $\phi^I$ $I =1, ...  ,6$ and four fermions
$ \chi_{\alpha i} $, $\chi_{\dot \alpha \bar i}$,  where $\alpha$
and $\dot \alpha$ are four dimensional chiral and anti-chiral spinor
indices respectively and $i =1,2,3,4$ is an index in the
{\bf 4} of SU(4) = SO(6) and $\bar i$ in the {$\bf \bar 4$}.
(The 4 is the spinor of SO(6)).
The theory has a global S0(6) symmetry. This symmetry does not
commute with the supercharges, since different components of the
multiplet have different SO(6) quantum numbers. In fact,
 the supercharges
are in the $\bf 4 $ and $\bf \bar 4 $ of SU(4). A symmetry that
does not commute with the supercharges is called an ``R'' symmetry.
Note that SU(4) is a chiral symmetry.

The Lagrangian is schematically of the form
\eqn\lagrang{
L = { 1 \over g^2 } Tr\left[
 F^2 +   (D\phi)^2 +
 \bar \chi \not D \chi +  \sum_{IJ} [\phi^I \phi^J]^2
+  \bar \chi \Gamma^I \phi^I \chi
\right] + \theta Tr[ F \wedge F]
}
It contains two parameters, the coupling constant and a theta angle.
The theory is scale invariant quantum mechanically. Namely
the beta function is zero  to all orders.
So it is also conformal invariant. The extra conformal symmetries
commuted with the 16 ordinary supersymmetries give 16 new supersymmetries.
In any conformal theory we have this doubling of supersymmetries.

The theory has an  S-duality under which
\eqn\coupl{
\tau_{YM} =  {\theta \over 2 \pi }+ i { 2 \pi \over g^2_{YM} }
}
transforms into $ -1/\tau$. This combines with shifts in the
$\theta $ angle into $SL(2,Z)$ acting on $\tau$ as it usually acts
on the upper half plane.
The 't Hooft coupling is $\lambda = g^2_{YM} N$.

\subsec{ IIB strings on $AdS_5 \times S^5 $. }

Suppose that the radius is large. We will later find under which
conditions  this is
true.
Then we are looking for a solution of the type IIB supergravity
equations of motion.
These equations follow form the action\foot{We have suppressed
the dependence of the action on the fields that are not
important for our purposes.}
\eqn\act{
S \sim  \int \sqrt{g} R  + F_5^2
}
plus the self duality constraint for the fiveform field strength,
 $F_5 = * F_5$, which has to be imposed by hand.
Due to the existence of $D3$ branes the flux of $F_5$ is quantized.
\eqn\fluxquant{
\int_{S^5} F_5 = N
}
Choosing a fiveform fieldstrength
 proportional to the volume
form on $S^5$ plus the volume form on $AdS_5$ we find that
$AdS_5 \times S^5$ is a solution.
The radius $R$ of the sphere and the radius of $AdS$ are
\eqn\radiusform{
R =  (4 \pi  g_s N)^{1/4} l_s \sim N^{1/4} l_{pl}
}
where $g_s$ is the string coupling and $2 \pi l_s^2$ is the
inverse of the string tension.
It is clear from \act\ that the radius in Planck units should
have this form since the $F_5^2$ term in the action scales like
$ N^2 $,
 while the first term in \act\ scales like $R^8$. The equations of motion
 will balance these two terms giving \radiusform .

It is also amusing to understand the energetics that gives rise
to the negative cosmological constant in $AdS_5$.  For this
purpose consider a compactification of the ten dimensional theory
on $S^5$ to five dimensions with a fiveform fieldstrength on an
$S^5$ of
radius $r$.
Then the five dimensional action is  schematically
\eqn\fivedimact{\eqalign{
S &= \int r^5 \sqrt{g_5} R^{(5)}  -
\sqrt{g_5} N^2 r^{-5} + \sqrt{g_5} r^3
 = \int \sqrt{g_E} R_{E} - V(r)
 \cr
& { \rm with} ~~~~~~ V(r) = r^{ - 25/3} ( N^2 r^{-5}   -  r^3)
}}
\ifig\figtwelve{ Effective potential after compactifying
ten dimensional supergravity on a five sphere with a flux of
the fiveform field strength.    }
{\epsfxsize 2.5 in\epsfbox{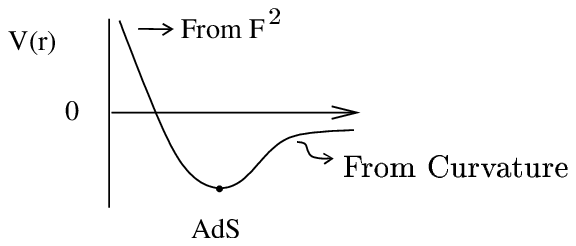}}
The second term
 comes from the flux and the third term come from the
curvature of $S^5$.
Note that the sign in Einstein's action is such
that an internal space with positive curvature gives rise
to a negative
contribution to the energy in the non-compact dimensions.
The potential should be computed in 5d Einstein frame.
This potential goes to zero as $r \to \infty $, which is  a general
feature of KK compactification.
One can also
see that there is a minimum that balances the two pieces with
an $r$ as in \radiusform .

\subsec{ ${\cal N} =4$ YM is the same as IIB on $AdS_5 \times S^5$}

Now we want to relate the two theories we have just talked about.
The general reason that they could be related is that in the 't Hooft
limit we expect strings. This string will move in a
space that has more than four dimensions.
 The field theory has 32 supersymmetries
which is the same as the number of supersymmetries
 of type  IIB string theory on this background.
In fact, the two supergroups are the same. So it is reasonable that
the two theories could be related.

There is an argument that relates these two theories which
relies on
looking at the near horizon geometry of D3 branes. The field theory
on $N$ D3 branes is ${\cal N} =4$ U(N) Yang Mills at low energies.
The near horizon geometry of $D3$ branes is $AdS_5 \times S^5$.
Since excitations that live near the horizon have very small
energies from the point of view of the outside observer we conclude
that at low energies only these excitations will survive.
So in the low energy limit we have two alternative descriptions
which should be equivalent \jm .

The coupling constants of YM are related to the string coupling
and vev of the RR scalar
\eqn\taures{
{ i \over g^2_{YM}} + {\theta \over 2 \pi} = \tau_{YM} =
\tau = { i \over g_s} + \chi
}
This notation emphasizes that both theories have an SL(2,Z) duality
symmetry.

The relation between the two theories is a ``duality". There is a
parameter, $g^2_{YM} N$, such that in the region where it is very
small one description (the Yang Mills one)
is weakly coupled and the other
(gravity) is strongly coupled, while the opposite is true
when this parameter is large.
Let us expand on this point.
 The gravity description is a good approximation
to string theory  if
the radius of the space is much larger than $l_s$, since
$l_s$ is the intrinsic size of the graviton. We see from \radiusform\
\taures\ that this happens when $ g^2_{YM} N \gg 1$.
It is good  that the two weakly coupled descriptions
are non-overlapping. Otherwise we would have
 blatant contradictions since the two
theories have rather different properties in their respectively
weakly coupled regimes. This fact also makes the conjecture hard
to disprove, or hard to prove. In this supersymmetric case there
are some quantities that are independent of the coupling
which  can be computed on both sides. Checking that these quantities
agree we have checks of the duality. For a more detailed
discussion of these checks see \aharonyetal .

Finally note that $\alpha'$ is not a parameter in string theory
so all physical quantities  depend only on the size of $AdS$
in string units. A useful way to think about it is to choose
units where the radius $R =1$,  then $\alpha' =
{ 1 \over \sqrt{ g^2_{YM} N}} $.
All gravity computations  depend only  $N$, but not on
$g^2_{YM} N$. The reason is that if we write the action in Einstein
frame then we have  an overall factor of $l_{pl}^{-8} \sim N^2$.
Then the $\alpha'$ expansion is an expansion in terms
of $ 1/\sqrt{g^2_{YM} N} $.

\newsec{ Establishing the dictionary}

\subsec{ Correlation functions}

We now consider correlation functions.
 We focus on  the Euclidean case.
The Euclidean CFT is dual to Euclidean $ADS$ which is the same as
hyperbolic space.
We can write coordinates as in \adsfive\ where now $dx^2$ denotes
the metric on $R^4$.

We can evaluate the gravity partition function as a function of
the boundary values of the fields. Since $AdS$ has a boundary
we need to specify the boundary conditions for the fields. The
value of the partition function depends on these boundary
conditions.
If the gravity theory is weakly coupled we can approximate this
by the value of the classical action
\eqn\partfn{
{\cal Z}_{bulk} \left[
 \phi(\vec x, z)|_{z=0} = \phi_0(\vec x) \right] \sim
 e^{ - N^2 S_{class}[\phi]  + o(\alpha') } \times ( {\rm Quantum Corrections} )
}
We emphasized the fact that the classical gravity computation will
always depend on $N$ through this overall factor in the action and
will be independent of $g^2 N$. Stringy corrections will correct
the gravity action into the classical string action, whose form
we do not  know, but we know that it will have an expansion in powers
of $\alpha' = { 1 \over \sqrt{g^2 N}}$. Some terms in this expansion
are known. For example, there is a
 well studied $R^4$ correction to the ten dimensional
action.

For each field in the 5 dimensional bulk, we have a corresponding
operator in the dual field theory.
In general, figuring out which operator corresponds to which field is
hard. But for some special operators it is easy due to their symmetries.
For example, the graviton  is associated to the
stress tensor operator in the boundary theory. Similarly the dilaton
is related to the Lagrangian of the theory,  since we saw that the
coupling is related to the dilaton and a change in coupling
adds an operator proportional to the Lagrangian.

The $AdS/CFT$ dictionary says that the quantity appearing in \partfn\
is equal to the generating function of the correlation functions of
the corresponding operators.
\eqn\fieldoperatorcor{
{\cal Z}_{bulk} \left[
 \phi(\vec x, z)|_{z=0} = \phi_0(\vec x) \right] =
 \langle e^{ \int d^4 x \phi_0(\vec x) {\cal O}(\vec x)} \rangle_{
 \rm Field ~Theory}
 }
 Note that $\phi_o(\vec x)$ is an arbitrary function specifying
the boundary values of the bulk field $\phi$.
 Taking derivatives with respect to $\phi_0$ and setting it to zero
 we obtain the correlation functions of the operator.

 So the final conclusion is that changes in the boundary conditions
 of $AdS$ correspond to changes in the Lagrangian of the field theory.
 Infinitesimal changes in the boundary condition correspond to the insertion
 of an operator.

 Now let us discuss this more explicitly.
 We consider the metric \adsfive . Consider a scalar field of mass $m$.
 Its action is
 \eqn\actionsal{
 S = N^2 \int { dx^4 dz \over z^5 } [
 z^2 ( \partial \phi)^2 + m^2 R^2 \phi^2 ]
 }
The action  contains higher order
 terms in $\phi$. For the moment we focus on the quadratic terms.

In order to evaluate the classical action on a classical solution
we need to solve the classical equations
\eqn\claseq{
z^3 \partial_z ( { 1 \over z^3} \partial_z \phi) - p^2 z^2 \phi
- m^2 R^2 \phi =0
}
We have used translation symmetry to go to Fourier space for the
$R^4$ part. This equation can be solved exactly in terms of
Bessel functions.
For the moment let us just understand the behavior of the solutions
near the boundary of $AdS$, at $z \sim 0$.
We then  look for solutions of \claseq\ in terms
of powers of the form $\phi \sim z^\alpha$, where $\alpha$
obeys the equation
\eqn\indicial{\eqalign{
 &\alpha (\alpha-4) - m^2 R^2 =0
\cr
 & \alpha_{\pm} = 2 \pm \sqrt{ 4 + m^2 R^2 }
 }}
The solution with $\alpha_-$ dominates near $z\to 0$.
The solution with $\alpha_+$ always decays when $z\to 0$.
We will impose the boundary condition on the dominating solution.
More precisely we choose a boundary condition of the form
\eqn\boundcond{
\phi(x, z)|_{z=\epsilon} = \epsilon^{\alpha_-} \phi^r_0( x)
}
In general we need to impose boundary conditions at $z= \epsilon$ and
then take $\epsilon \to 0$ at the end of the computation.
We see that if we keep $\phi^r_0(x)$ fixed as we take $\epsilon \to 0$
the solution in the bulk, at some fixed $z$ will have a finite
limit. So we call this $\phi_0^r(x)$ the ``renormalized'' boundary
condition. This is related to the fact that
in the field theory we need to renormalize the operator.

Under  a rescaling of coordinates in the field theory, which
in AdS is the isometry $x \to \lambda x, ~ z \to \lambda z $,
the original field $\phi$ does not get scaled, but due to the
$\epsilon$ factor in \boundcond\ we see that
$\phi^r_0$ has dimension $\alpha_-$.
Since we interpret the resulting gravity formulas
through \fieldoperatorcor , where on the right hand side we
have $\phi_0^r(x)$, we conclude that the dimension of the
corresponding
operator is
\eqn\dimension{\eqalign{
\Delta =& 4 - \alpha_- = \alpha_+
\cr
\Delta =& 2 + \sqrt{ 4 + (mR)^2}
}}

Note that for the dilaton, which has $m=0$, we get the
correct dimension, $\Delta =4$.
Since
the theory is exactly conformal for all values of $g_{YM}$ the
operator that changes infinitesimally the Yang Mills coupling
should have dimension four.
Similarly, the fact that the graviton is massless is related
to the fact that the stress tensor has dimension four. This last
fact  does not depend on supersymmetry and is very general, valid for
any local conformal QFT.

Note that in order to make the operator to field correspondence
it is necessary to Kaluza-Klein reduce all fields to five dimensions.
So when we talked about the dilaton, we were referring to the
$l=0$ mode on the five-sphere which gives rise to a massless field on
$AdS$. A mode with angular momentum $l$ on the five-sphere has mass
given by $m^2 = l(l+4)/R^2 $ which leads to
\eqn\dimdil{
\Delta = 4 + l ~.
}
The corresponding operators are roughly of the form
\refs{\gkp,\wittenhol}
\eqn\operdil{
 Tr[ F^2 \phi^{(I_1 } \cdots \phi^{I_{l}) } ]
 }
 where the indices $I_i$ are taken in a symmetric traceless combination
 which corresponds to the representations of the spherical harmonics.
 In \operdil\ one should order the fields appropriately and it
 is also necessary to introduce the fermionic fields
\ref\KlebanovXV{
I.~R.~Klebanov, W.~I.~Taylor and M.~Van Raamsdonk,
Nucl.\ Phys.\ B {\bf 560}, 207 (1999)
[arXiv:hep-th/9905174].
}.
 The dimension of \operdil\ can
 be computed easily at weak coupling, we just sum the dimensions
 of the individual fields to get $\Delta = 4 +l$. We see that
 we get the same as the strong coupling result \dimdil .
 The reason is that the operators \operdil\ are in protected multiplets.
 A protected multiplet is a multiplet of supersymmetry that is
 smaller than the generic multiplet. Such multiplets have dimensions
 which are given in terms of their SO(6) quantum numbers.
Therefore, such operators  cannot have
 coupling dependent anomalous dimensions.

 It turns out that in the case of ${\cal N} =4 $ YM all
 the operators that are in protected multiplets correspond to
 all the KK modes of the gravity fields. This provides a nice
 match since we have the same number of protected states on both
 sides. This matching goes beyond the matching of symmetries of the
 two theories. In fact, we could have obtained  extra protected
 operators in the YM theory, for example. This would have killed
 the conjecture since we do not have any other light states in
 supergravity. As an example, note that if we changed the gauge
 group to SO(N) instead of U(N) then we have we do not have
 all the operators \operdil , we only have the ones with even
 $l$. In fact, the SO(N) theory corresponds to an orientifold of
 $AdS_5 \times S^5$ that maps antipodal points on $S^5$
\ref\WittenXY{
E.~Witten,
JHEP {\bf 9807}, 006 (1998)
[arXiv:hep-th/9805112].
} \foot{ There
 are four  distinct orientifolds, one gives SO(2N), one gives
 SO(2N+1)  and the other two give two versions of the
 Sp(N) theory \WittenXY.}.

Note that not all operators are protected. For example, the operator
$tr[\phi^I \phi^I] $ has dimension two at weak coupling but there
is no corresponding operator with dimension two at large coupling.
What is its dimension at strong coupling?. As we discussed,
all supergravity modes have dimensions that remain fixed as
 $g^2 N \to \infty $. But a string theory also contains massive string
 states, with masses $m \sim 1/l_s$, which according to \dimension\
 correspond to operators of dimension $\delta \sim R/l_s \sim ( g^2_{YM}
 N )^{1/4} $. So the dimension of this operator should at least be
 of this order of magnitude at strong 't Hooft coupling.
In fact, the Yang Mills theory contains many operators with higher
spin, like $tr[ \phi^I \partial_{(\mu_i } \cdots \partial_{\mu_s)} \phi^i]
$. These operators have dimension $2 +s $ at weak coupling but they
should also get dimensions of order $(g^2_{YM} N)^{1/4}$ at strong
coupling since the gravity theory only contains fields of spin
less than two. In any theory that has a weakly coupled
gravity dual, with a radius of curvature much bigger than the
string scale,
 operators with higher spin should have large dimensions.

 The problem of solving the Laplace  equation with fixed boundary
 condition is rather familiar from electrostatics. It is
 useful to introduce the bulk to boundary propagator, which is
 the solution of the problem where we put a boundary condition
 that is a delta function at a point on the boundary.
 In this context this bulk to boundary propagator is \wittenhol\
 \eqn\bulkbdy{
 G_{\Delta}(z, \vec x, \vec x') = { z^\Delta \over
 [ (\vec x - \vec x' )^2 + z^2 ]^\Delta }
 }
 We can use this to compute connected correlation functions.
 For example if we wanted to compute the connected three point
 function then we would have to include possible cubic terms in the
 action for the scalar field
 \eqn\actcub{
 S = N^2 \int (\nabla \phi)^2 + m^2R^2  \phi^2 + \phi^3 + \cdots
 }
\ifig\figthirteen{ Diagram contributing to a three point function.
The vertex is a $\phi^3$ interaction in the bulk theory. The lines
going to the boundary are bulk to boundary propagators. }
{\epsfxsize .75 in\epsfbox{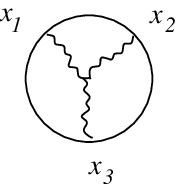}}
 We need to evaluate the diagram in  \figthirteen .
This  leads to the following expression for the three point
function
\eqn\express{
\int { d^4 x dz \over z^5} G_{\Delta_1}(z,x,x_1) G_{\Delta_2}(z,x,x_2)
G_{\Delta_3}(z,x,x_3)
}
with $G$ as in \bulkbdy . This integral gives, of course, the
$x$ dependence for a  three point function that we expect on
the basis of conformal invariance.
Remember that in a conformal invariant theory the two and three
point function are given by
\eqn\two{
\langle {\cal O } {\cal O} \rangle = { 1 \over |x_1 - x_2|^{ 2 \Delta} }
}
\eqn\three{
\langle {\cal O}_1 {\cal O }_2 {\cal O }_3 \rangle =
{ C_{123} \over | x_1 - x_2|^{ \Delta_1 + \Delta_2 - \Delta_3}
| x_3 - x_2|^{ \Delta_3 + \Delta_2 - \Delta_1}
| x_1 - x_3|^{ \Delta_1 + \Delta_3 - \Delta_2}
}}
Indeed the integral \express\ gives \three , with
$\Delta_i = \Delta$ \ref\FreedmanTZ{
D.~Z.~Freedman, S.~D.~Mathur, A.~Matusis and L.~Rastelli,
Nucl.\ Phys.\ B {\bf 546}, 96 (1999)
[arXiv:hep-th/9804058].
}.

Normally we can normalize the two point functions to one.
There are some special two point functions for which the normalization
is unambiguously defined. An example is the two point function of the
stress tensor. This two point function is proportional to $N^2$ with
a coefficient that agrees in gravity and field theory. In fact this
coefficient does not depend on the coupling due to a supersymmetry
argument \ref\GubserSE{
S.~S.~Gubser and I.~R.~Klebanov,
Phys.\ Lett.\ B {\bf 413}, 41 (1997)
[arXiv:hep-th/9708005].
}.

\subsec{Various remarks}

Note that when we solve the equation \claseq\ we need to impose
two boundary conditions since it is a second order equation.
One is the one we discussed so far. The second is that it should vanish
as $z \to \infty $. This corresponds to the statement that there is
nothing special happening at infinity. In Euclidean $AdS$ the
 point $z = \infty $
is actually a point on the boundary.
In general,
 when we solve the equation we need to impose the condition
that the solution is not singular in the interior, this
gives us the second boundary condition.

Note also that the renormalization  in \boundcond\
is independent of $p$ (or $x$). This is  related to the fact
that the dual theory is an ordinary local quantum field theory where
the renormalization of an operator does not depend on the momentum.
In other theories such as in linear dilaton backgrounds, or in
non-commutative field theories, this is no longer true
\ref\PeetWN{
A.~W.~Peet and J.~Polchinski,
Phys.\ Rev.\ D {\bf 59}, 065011 (1999)
[arXiv:hep-th/9809022].
}
.

 The fact that the theory on the boundary is local
implies that the theory in the bulk should contain
gravity. By the word ``local'' we mean that the theory
contains a stress tensor. This in turn means that
the theory can be  defined on any manifold.

It is possible to have fields with negative mass squared
in $AdS_5$ as long
as the mass obeys
\eqn\bfreed{
 m^2 R^2 \geq -4
 }
These tachyons do not lead to instabilities and actually appear
in $AdS_5 \times  S^5$. They do not lead to instabilities because
normalizable  wavefunctions in global AdS,
have positive energy. This is due to
the fact that the wavefunction
has to decrease as $\rho \to \infty $, which
 implies that it should have
some kinetic energy which overwhelms the negative potential energy.
These fields correspond to relevant operators. In fact, one can
see that $\alpha_-$ is positive.  This means that the perturbation
they induce, which goes as $\phi \sim z^{\alpha_-}$,
 decreases as $z \to 0$, which is the UV of the field
theory\foot{For some of these fields one sometimes effectively imposes the
boundary conditions on the solution that decreases faster so that
their dimension is equal to $\alpha_-$ instead of \dimension
\ref\KlebanovTB{
I.~R.~Klebanov and E.~Witten,
Nucl.\ Phys.\ B {\bf 556}, 89 (1999)
[arXiv:hep-th/9905104].
}.
This change in the boundary condition is
 not needed for ${\cal N}=4 $ YM but it is  needed for
some other theories.}.
Note also that a field of zero mass corresponds to a marginal deformation
and in this case $\alpha_-=0$ so that the perturbation produced by
the operator is independent of $z$. Finally, if $m>0$, the perturbation
increases as $z \to 0$. This corresponds to an irrelevant operator.

The conformal group has a very special representation at $\Delta =1$.
Unitarity implies that the corresponding operator corresponds
to a free field in $R^4$. These operators arise in the $U(1)$ factor
of the field theory. This is the $U(1)$ in $U(N)$, the operator
is $Tr[ \phi^I] $. These representations
of the conformal group are called singletons.
They are special because
they do not correspond to an ordinary field propagating in the bulk.
They should have no local degrees of freedom in the bulk, only on the
boundary.
Another example is the operator $Tr[ F_{\mu \nu}]$.
 It corresponds in
the bulk to the $l=0$ modes of the RR and NS
$B_{\mu \nu}$ fields. These fields have a long distance
action governed by
\eqn\actform{
 N \int B^{NS} \wedge d B^{RR}
 }
This is purely topological and gives rise
 to a field on the boundary if we put local boundary conditions
\ref\AharonyQU{
O.~Aharony and E.~Witten,
JHEP {\bf 9811}, 018 (1998)
[arXiv:hep-th/9807205].
}
\ref\MaldacenaSS{
J.~M.~Maldacena, G.~W.~Moore and N.~Seiberg,
JHEP {\bf 0110}, 005 (2001)
[arXiv:hep-th/0108152].
}.

 \subsec{ Physics of the warp factor}

Let us try to understand the role of the radial
coordinate $z$ in the bulk theory.
It is very important that the metric contains a redshift factor,
or
warp\foot{A redshift factor multiplies $dt$ and a warp factor is basically
a redshift factor which multiplies several spacetime coordinates
so that we have Poincare invariance.} factor
that multiplies the metric
\eqn\metricw{
ds^2 = w(z)^2 (  dx^2 + dz^2)  ~,~~~~~~w(z) = { 1 \over z}
}
 The distances and times in the
field theory are measured with the coordinates $x$.
On the other hand proper times and proper distances in the bulk
are related to the field theory coordinates by the warp factor
$w \sim 1/z$.
So a given object in the bulk, such as a massive string state,
corresponds to field theory configurations
 of different size and energy depending
on the value of the $z$ coordinate of
 the bulk object.
We have
\eqn\ener{
E_{FT} = w(z) E_{proper}
}
\eqn\sizef{
({\rm size})_{FT} = { 1\over w(z)} ( {\rm proper~size})
}
where the $_{FT}$ subindex indicates a quantity in the field theory.

So we see that as we go to $z \to 0$ we have very small sizes in the
field theory and very high energies. This is the UV of the field theory.
Notice that going to small $z$ corresponds to going to large distances
from a point in the interior.
In fact UV divergences in the field theory are related to
IR divergences in the gravity theory \ref\SusskindDQ{
L.~Susskind and E.~Witten,
arXiv:hep-th/9805114.
}.
The fact that short distances in the field theory correspond to
long distances in the gravity description is
  called the ``IR/UV correspondence''.
\ifig\figfifteen{ The same bulk object at different $z$ positions
corresponds to an object in the CFT with different sizes. }
{\epsfxsize 4 in\epsfbox{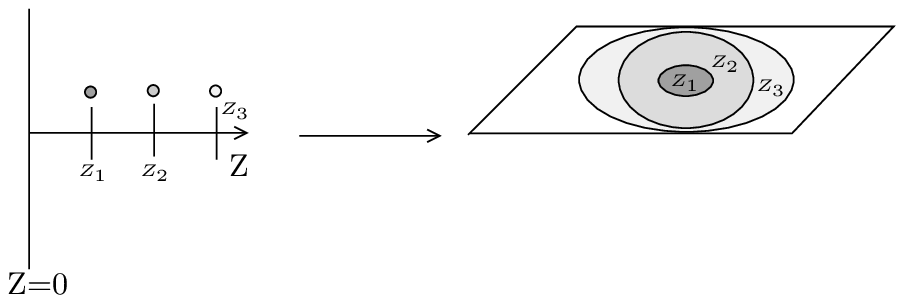}}
The physics of the warp factor is also responsible for pointlike,
or partonic,
behavior of scattering amplitudes \ref\PolchinskiTT{
J.~Polchinski and M.~J.~Strassler,
Phys.\ Rev.\ Lett.\  {\bf 88}, 031601 (2002)
[arXiv:hep-th/0109174]; JHEP {\bf 0305}, 012 (2003)
[arXiv:hep-th/0209211].JHEP {\bf 0305}, 012 (2003)
[arXiv:hep-th/0209211].
}.

\newsec{Thermal aspects}

Consider black holes in $AdS$. The five sphere will not
play an important role so we will not write it explicitly.
In Poincare coordinates the
simplest black hole is a black brane which is
 translation invariant along
the three spatial directions of the boundary. The metric has the
form
\eqn\blmet{
ds^2 = R^2{ 1 \over z^2}  \left[
 - ( 1 - { z^4 \over z_0^4} )dt^2 + d{ \vec x}^2 + { dz^2 \over
(1 - { z^4 \over z_0^4} ) }   \right]
}
The temperature is easily determined by going to Euclidean time
and choosing a periodicity for Euclidean time to that there is
no singularity at $z = z_0$. We see that this gives
$\beta = \pi z_0 $.

We can compute the entropy using the Bekenstein Hawking formula
$S = {({\rm Area}) \over 4 G_N} $ and we get the entropy per unit
volume \ref\GubserDE{
S.~S.~Gubser, I.~R.~Klebanov and A.~W.~Peet,
Phys.\ Rev.\ D {\bf 54}, 3915 (1996)
[arXiv:hep-th/9602135].
}
\eqn\entropy{
 {S \over V} = { \pi^2 \over 2}  N^2 T^3
}
The weakly coupled field theory has entropy per unit volume
\eqn\entrweak{
{ S \over V} = { 4 \over 3 } { \pi^2 \over 2} N^2 T^3
}

Note that the temperature dependence is determined by conformal
invariance. Note also that these black branes have positive
specific heat.

The $N$ dependence (for fixed $g^2N$) is as we expect from large
$N$ counting.
In principle one might have expected extra $g^2N$ dependence
at strong coupling, but we see that at very strong coupling the
answer \entropy\ becomes independent of $g^2N$. Remember that we
said that all gravity computations are independent of $g^2N$.
We see that the numerical coefficient is different. This is not surprising
since this is not a protected computation and the answer could
depend on the coupling $g^2 N$. In fact the leading $g^2N $
correction to
\entrweak\ has been computed \ref\FotopoulosES{
A.~Fotopoulos and T.~R.~Taylor,
Phys.\ Rev.\ D {\bf 59}, 061701 (1999)
[arXiv:hep-th/9811224].
}. The first $\alpha'$ correction
to the gravity result \entropy\ has also been computed, it comes
from an $R^4$ term and is proportional to $ {\alpha'}^3 \sim
(g^2 N)^{-3/2} $\ref\GubserNZ{
S.~S.~Gubser, I.~R.~Klebanov and A.~A.~Tseytlin,
Nucl.\ Phys.\ B {\bf 534}, 202 (1998)
[arXiv:hep-th/9805156].
}. Both corrections go in the right direction but
nobody has computed yet the whole interpolating curve.
In other words, we expect a behavior of the form
\ifig\figsixteen{ Form of the Yang Mills free energy as a
function of the 't Hooft coupling in the large $N$ limit. The
doted line is a naive interpolation between the weakly coupled
and the strongly coupled results. }
{\epsfxsize 2.5 in\epsfbox{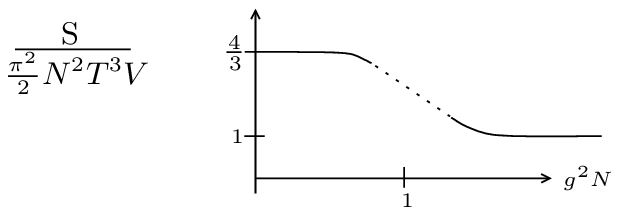}}

If we consider $AdS$ in global coordinates then we can
consider solutions that correspond to localized black holes
sitting at the center. If the Schwarzschild radius of these
black holes is very large, larger than the radius of $AdS$, then
they behave as the black branes we discussed above, but with
the $R^3$ directions replaced by $S^3$. On the other hand
if their radius is much smaller than the radius of $AdS$ they
behave more as ordinary Schwarzschild black holes in flat space.
They have negative specific heat and are unstable. It would be
nice to find their  precise description in the gauge theory.

\subsec{ Wilson Loops}

A very interesting operator in gauge theories is
\eqn\wloop{
W({\cal C}) = Tr[ P e^{\oint_{\cal C} A} ]
}
where ${\cal C}$ is a closed curve in $R^4$. The trace can be
taken in any representation, but we will take it in the fundamental
representation.
This operator represents the Yang Mills contribution to the
propagation of a heavy quark in the fundamental representation.

The general large $N$ counting arguments that we reviewed above
tell us that a quark  in the fundamental will have a
string ending on it.
So we expect to have a string worldsheet
 with a boundary along the contour
${\cal C} $. As usual this worldsheet lives in 5 dimensions and
ends on the boundary along ${\cal C}$.

\ifig\figseventeen{String worldsheet ending on the contour
${\cal C}$ corresponding to the trajectory of a heavy external
quark. }
{\epsfxsize 1.5 in\epsfbox{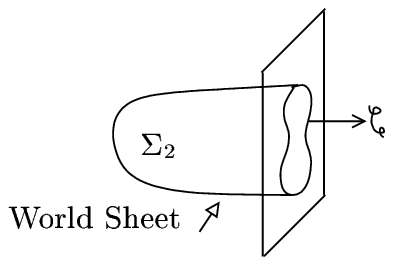}}

In $AdS_5 \times S^5$ we also have to specify at what point of
$S^5$ the string is sitting when it approaches the boundary.
In fact the operator that corresponds to a string at a specific
point on $S^5$ has the form \ref\MaldacenaIM{
J.~M.~Maldacena,
Phys.\ Rev.\ Lett.\  {\bf 80}, 4859 (1998)
[arXiv:hep-th/9803002].
} \ref\ReyIK{
S.~J.~Rey and J.~T.~Yee,
Eur.\ Phys.\ J.\ C {\bf 22}, 379 (2001)
[arXiv:hep-th/9803001].
}
\eqn\correctop{
W = tr[ P e^{ \oint A_\mu dx^\mu  + |dx| \phi^I(x) \theta^I}]
}
where $\phi^I$ are the scalar fields of ${\cal N}=4$ YM and
$\theta $ is a unit vector, $\theta^2=1$, which specifies a point
on $S^5$.

For large $g^2N$ the leading contribution to the expectation
value of the Wilson operator is of the form
\eqn\formw{
\langle W \rangle \sim  e^{ - T ({\rm Area})} \sim e^{-
 \sqrt{g^2 N} (
{\rm Area})_{R=1} }
}
Note that we have a factor of $1/\alpha'$ in the exponent. This
has explicit $\sqrt{g^2 N}$ dependence because it involves a
string, it goes beyond the supergravity fields.

The area in \formw\ is the proper area in the five (or ten) dimensional
space. It is infinite
since we have already seen that the proper distance to the boundary
is infinite, as is the area.
We need to regularize the expression and
compute the area up to $z = \epsilon $. Then we find that the
area goes as
\eqn\reguls{
{\rm Area} = { ({\rm length}) \over \epsilon} + A_r + o(\epsilon)
}
where $A_r$ is finite and can be called the renormalized area. The
divergent term is proportional to the length of the contour
${\cal C}$. This just renormalizes the mass of the external quark.

The simplest example is a circular contour  where we get
that $A_r$ is a negative constant. The
 result is independent of the size of the
circle due to conformal invariance.
In fact, for a circular contour there is a trick
that enables us
to do the exact computation \ref\EricksonAF{
J.~K.~Erickson, G.~W.~Semenoff and K.~Zarembo,
Nucl.\ Phys.\ B {\bf 582}, 155 (2000)
[arXiv:hep-th/0003055].
} \ref\DrukkerRR{
N.~Drukker and D.~J.~Gross,
J.\ Math.\ Phys.\  {\bf 42}, 2896 (2001)
[arXiv:hep-th/0010274].
}. This trick is based on the
observation that the straight line would give zero since it
corresponds to a BPS state. But the circle is related to the
straight line by a conformal transformation. The mapping of the point
at infinity is subtle so all the contribution comes from an
anomaly in the transformation \DrukkerRR .
{}From this exact answer one can see that in the limit of
large $g^2 N$ we get the right answer.

Another simple example is the computation of a quark anti-quark
potential which can be obtained by considering a rectangular
contour. This gives
\eqn\potent{
V =  - c_1 { \sqrt{g^2_{YM} N} \over L }
}
where $c_1$ is a numerical  constant.
The dependence on $L$ follows from conformal invariance.
The weak coupling result is
\eqn\weakcoup{
V = - c_2 { g^2_{YM} N \over L }
}
where $c_2$ is a numerical constant.
Even though the Yang Mills theory has a string description the
theory is not confining. (We will later see examples of confining
theories). The reason this happens is because the string moves
in five dimensions. The Wilson loop obeys the area law, but in
5-dimensions!

\ifig\figeighteen{ String configurations relevant for computing
the quark-anti-quark potential. As we separate the quarks the
string moves into a region with smaller warp factor, which
makes
the renormalized area smaller.  }
{\epsfxsize 1.5 in\epsfbox{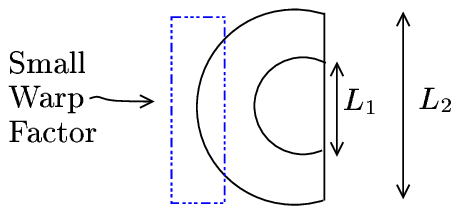}}
The potential decreases when we increase $L$ because the string
moves into the region of large $z$ where the warp factor is
smaller, so that its proper renormalized area is smaller,
see  \figeighteen .

\newsec{ Confining theories }

Here we will consider the simplest example of a confining theory
\ref\WittenZW{
E.~Witten,
Adv.\ Theor.\ Math.\ Phys.\  {\bf 2}, 505 (1998)
[arXiv:hep-th/9803131].
}.
We start with
 3+1 dimensional ${\cal N} = 4 $ Yang Mills  and compactify it
on a circle of radius $r_y$ down to $2+1$ dimensions.
So the theory is on the space $R^{2+1} \times S^1$.
If we choose antiperiodic boundary conditions on the circle we will
break supersymmetry. From the 2+1 dimensional point of view the
fermions will be massive.
Through quantum corrections, which are large if $g^2 N$ is large,
the bosons will also get a mass.
So the only massless fields will be the gauge bosons. So we
have a pure Yang Mills theory in 2+1 dimensions at low
energies. This theory is confining.

The supergravity description of this theory can be obtained in a
simple way from our previous solution \blmet .
This Euclidean black hole  corresponds to the Euclidean
field theory with one direction compact with an antiperiodic
boundary condition for the fermions in this compact direction.
We can then go back to Lorentzian signature by taking time
to be one of the non-compact coordinates. The resulting metric
is a double wick rotation of \blmet
\eqn\confmet{
ds^2 = R^2 {1\over z^2} \left[
 -dt^2 + dx_1^2 + dx_2^2  +
 ( 1 - { z^4 \over z_0^4} )dy^2  + { dz^2 \over
(1 - { z^4 \over z_0^4} ) }  \right]
}
where
we have suppressed the five sphere which will not play a role in
our discussion.
Now $y = y + 2 \pi r_y$ and $z_0 = 2 r_y$ is determined by demanding
that \confmet\ is nonsingular. Note that the topology of the boundary
of \confmet\ is $S^1 \times R^{2+1} $ and the topology of the full
space is $D^2 \times R^{2+1}$. So the circle is contractible in
the full space. The radial direction $z$ together with the circle
$y$ have the topology of a disk.

The crucial property of \confmet\ is that the space terminates
in the large $z$ direction and the warp factor is bounded below,
$ w(z) \geq w(z_0) \sim 1/z_0 $. Notice that the metric
\confmet\ does not have a horizon.
A particle moving in the metric \confmet\ feels a gravitational
force towards $z = z_0$. So the lowest energy states live at
$z = z_0$, the region of  space where the warp factor is smallest.
In fact, once we go to the quantum theory we expect that even
massless particles get a non-zero mass due to the fact that
they are moving in this gravitational potential well
 which forces the
wavefunction to vary in the $z$ direction since a  normalizable
wavefunction should go to zero at the boundary, $z \to 0$.
In fact, all excitations on this geometry have positive mass from the
$2+1 $ dimensional point of view.

In order to find the $1+2$ dimensional particle spectrum we
start from the five dimensional fields, let us say a scalar field
of mass $m$. We then solve the classical equation
\eqn\classeq{
 |g^{00}| \omega^2 \phi + { 1\over \sqrt{g} } \partial_z
 ( \sqrt{g} g^{zz} \partial_z \phi ) - m^2 R^2 \phi =0
}
For simplicity we assumed that $\phi$ is independent of $y$.
This equation should be solved with the boundary condition
that it vanishes as $z \to 0$ and that it is regular at
$z = z_0$, which  means $\partial_z \phi|_{z_0}  =0$.
So we really have an eigenvalue equation, i.e. the equation
with these boundary conditions can have a non-zero solution
only for special values of $\omega^2$. These special values
are the masses of the particles in $2+1 $ dimensions.
By multiplying by $\phi$ and a suitable power of $z$ and integrating
\classeq\ it is possible to see that the $w^2$ eigenvalues
are strictly positive. This can also be shown for the tachyons
that we have in $AdS$.

So we see that the theory has a mass gap. i.e. all excitations
have positive mass. This is a property we expect in confining
theories.
Note that these particles are glueballs from the point of view
of the boundary theory. These have masses of the order of the mass
gap, i.e. of order
\eqn\massgap{
 M \sim 1/z_0 \sim 1/r_y
}

The fact that the warp factor is bounded below also implies that
the Wilson loop will now lead to an area law in the boundary theory.
What happens is that the string will sit at $z= z_0$ and it will have
a finite tension there. I cannot move to a region where the
warp factor is smaller.

\ifig\fignineteen{ String configuration relevant for the computation
of the Wilson loop in a confining theory. Since the
warp factor is bounded below we have a finite string tension
from the boundary theory point of view. }
{\epsfxsize 1.5 in\epsfbox{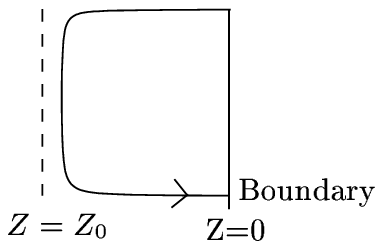}}

The string tension from the point of view of the boundary
theory is of order
\eqn\strtens{
T = { 1 \over \alpha'} w(z_0)^2 \sim { \sqrt{ g^2 N} \over r_y^2 }
}

\subsec{ Confinement-deconfinement transition}

Consider now this $2+1$ theory at finite temperature.
In the Euclidean description we compactify time with period $\beta$.
Going back to the 4d theory we now see that the theory is on
the space $R^2 \times S^1_\beta \times S^1_{r_y}$, so that
we have two circles. Then there are two Euclidean gravity solutions.
We can take the above solution \confmet\ and compactify Euclidean
time. Or we could take the same type of solution but exchanging
the two circles and adjusting $z_0$ appropriately. This gives
two different ways of matching a five dimensional geometry
to our four dimensional boundary. In the first solution the
$y$ circle is contractible but the Euclidean time circle is not.
The opposite is true in the other solution.
The
first solution
 corresponds to just considering the original 5d space \confmet\
at finite temperature, so we will excite thermally some of the particles
we had discussed above. This is the solution that has lowest free energy
at low temperatures. On the other hand the solution where the
Euclidean time direction is contractible corresponds to having a
black brane of the type we discussed above. This solution has
lower free energy at high temperatures. There is a critical
temperature at which they have the same free energy. At this
temperature we have a first order phase transition. In this
simple case the symmetries of the problem imply that
the critical temperature is at
 $\beta_c = 2 \pi r_y$.
 This is the
confinement-deconfinement phase transition. It is of first order
because the entropy changes dramatically. In the low temperature
phase we have an entropy independent of $N$, we only excite
color neutral glueballs whose spectrum is  $N$ independent.
The high temperature phase has an entropy given by the Hawking
Bekenstein formula which is proportional to $N^2$. We interpret this
as saying that now the gluons can move independently.

Note that in order to get this piece of physics correctly it was
important to sum over all geometries which are asymptotic to
a given boundary. This is a general principle in the duality.
The choice of theory, i.e. the choice of Lagrangian, only fixes
the boundary conditions of the gravity solution. Then we have
to sum over all geometries with these boundary conditions. The
leading contribution comes, of course, from classical solutions. In
some situations there are several classical solutions. We should
sum over all of them. The one with the lowest action will contribute
the most. We have phase transitions when one dominates over the other.
These are generically large $N$ phase transitions, which
can happen even in
finite volume.  Whether they
are bona-fide phase transitions at finite $N$
 has to be thought about more carefully.

Is this really a solution for the pure bosonic Yang Mills theory
in $2+1$ dimensions?. Not really. The reason is the following.
Bosonic Yang Mills theory in 2+1 has a dimensionfull coupling
$g^2_3 N$. This
theory is expected to have a confinement scale of order
$\Lambda \sim g^2_3 N$. On the other hand if this theory arises as the
low energy limit of some other theory, then this description in
terms of $2+1$ Yang Mills is quite reasonable if the scale of the
new physics, let us call it $\Lambda_{UV} $ is
\eqn\condition{
\Lambda_{UV} \gg \Lambda = g_3^2 N
}
This means  that the theory is weakly coupled
for energies in between $\Lambda$ and $\Lambda_{UV}$.
In our case the three dimensional coupling that we obtain from
dimensionally reducing the 4d theory is
\eqn\threed{
g^2_3 N = { g^2 N \over r_y}
}
while the scale at which we have new physics is $1/r_y \sim \Lambda_{UV}$.
We see that we can never obey \condition\ if we want to
trust supergravity, which requires $g^2 N \gg 1$.

So the theory that supergravity is describing is a confining theory
but it is not pure $2+1 $ Yang Mills.

There are many examples of confining theories in four dimensions. Some
are supersymmetric ${\cal N} =1$ theories that are confining
\ref\KlebanovHB{
I.~R.~Klebanov and M.~J.~Strassler,
JHEP {\bf 0008}, 052 (2000)
[arXiv:hep-th/0007191].
}. See M. Strassler's lectures.

Theories that have large radius supergravity duals are special.
For example glueballs of spin
greater than two will be much more massive than the
 lightest glueballs.
Similarly the string tension will be much larger than the mass
scale set by the lightest glueball, as we can see from
\massgap \strtens. It is believed that in large $N$ bosonic
 Yang Mills in four dimensions
these two scales are of the same
order. This would imply that the theory does not have a large
radius gravity
dual. We expect a dual description in terms of strings moving
on a space whose curvature is of the order of the string scale.

\subsec{ Remarks about more general field theories}

Many theories have gravity duals and  they do not need to
be conformal.
For local quantum field theories
  the corresponding gravity solution has the
feature that the warp factor becomes large towards the boundary.
If the theory is free in the UV, then the geometry becomes singular
as  the boundary is approached, the radius of curvature in string
units becomes very small.

If a CFT is deformed by adding some relevant operators to the Lagrangian
then there can be interesting effects in the IR. A simple
possibility is that the theory flows in the IR to another CFT.
In some cases people have found interpolating solutions that start as
one $AdS$ space near the boundary and end as another $AdS$ space in the IR
region \ref\FreedmanGP{
D.~Z.~Freedman, S.~S.~Gubser, K.~Pilch and N.~P.~Warner,
and a  c-theorem,''
Adv.\ Theor.\ Math.\ Phys.\  {\bf 3}, 363 (1999)
[arXiv:hep-th/9904017].
}.
One can prove that the supergravity equations imply that the radius
of curvature in 5d Planck units of the $AdS$ space decreases as you
flow to the IR. This is a supergravity $c$ theorem \FreedmanGP
\ref\AnselmiFU{
D.~Anselmi, L.~Girardello, M.~Porrati and A.~Zaffaroni,
Phys.\ Lett.\ B {\bf 481}, 346 (2000)
[arXiv:hep-th/0002066].
}.

If the theory is confining, then one typically finds that the
space ends as you go into the interior and that the warp factor is
bounded below. How precisely it ends depends on the theory under
consideration. It can end in a purely geometric way,
 as we saw above,
or there can be some branes \ref\PolchinskiUF{
J.~Polchinski and M.~J.~Strassler,
arXiv:hep-th/0003136.
}.

\subsec{ D-branes in the bulk}

We have a $U(N)$ theory on the boundary. But  the
$U(1)$ is a free factor. The physics in the interior of the
bulk is really described by the SU(N) piece.
In SU(N) $N$ quarks can combine into a neutral object.
In the bulk we can have $N$ fundamental strings that end on
a D5 brane that is wrapping the $S^5$ \WittenXY.

Normally we cannot have a fundamental string ending on a brane with
compact volume since the endpoints of fundamental strings act
as electric charges for the $U(1)$ gauge field living on the D-brane
worldvolume. Of course we could have a string ending and one
``departing", the orientation is important.

However, there is an interesting coupling on the D5 worldvolume of
the form
\eqn\coupl{
S_{brane} \sim \int d^6 x A \wedge F_5 \sim N \int dt A_0
}
where $A$ is the worldvolume gauge field. This is saying that the
5-form field strength induces $N$ units of background
electric charge on the 5-brane. This can be cancelled by $N$ strings
that end on it.

In other theories, which have dynamical quarks,
 we can have baryons as states in the theory.

Note that if we add flavors to the field theory we will get
D-branes that are extended along all 5 dimensions of $AdS$. The open
strings living on them are the mesons. The gauge fields living on them
are associated to flavor symmetries.

Note that gauge symmetries in the bulk correspond to global symmetries
in the boundary theory. In a gravitational theory only
gauged symmetries have associated conserved charges.

\newsec{ The plane wave limit of AdS/CFT}

Other reviews of this subject can be found in \ppwavereview .

\subsec{ Plane waves}

Plane waves are spacetimes of the form
\eqn\planew{
ds^2 = -2 dx^+ dx^-  - A_{ij}(x^+) y^i y^j (dx^+)^2 + d\vec y^2
}
The index $i=1, \cdots, D-2$.
These spacetimes have many isometries. One is obvious,
$ \partial_- $. To describe the other ones let us assume for the
moment that there is only one coordinate $y$ and call $A_{11} = \mu^2$.
Then the other isometries have the form
\eqn\isometr{
 a \equiv   \zeta(x^+) \partial_y  + \dot \zeta(x^+) y \partial_-
 }
where $\zeta(x^+)$ is a complex solution of the equation
 \eqn\zetaeq{
  \ddot \zeta + \mu^2(x^+) \zeta = 0
  }
  In the particular case that $\mu >0$ is constant a simple solution is
  $\zeta = e^{-i \mu x^+} $.
  Since $\zeta$ is complex there are two isometries in \isometr\ the
  other is the complex conjugate $a^\dagger =  a^*$.
We can normalize the solution of \zetaeq\ so that the commutator
of these two isometries is
\eqn\commut{
[ a , a^\dagger ] = i \partial_- = - p_-
}
If we have $n$ $y$ coordinates, then we have $n$
 $a, ~a^\dagger$ pairs.

The fact that the space has many isometries enables us to find the
solution of the geodesic equation and also enables us to separate the
Klein Gordon equation.

Let us see this more explicitly. The action for a particle is
\eqn\act{\eqalign{
S =&  {1\over 2} \int d\tau [e^{-1} \dot X^\mu \dot X^\nu g_{\mu\nu}
 - m^2 e ] \cr
S =& { 1 \over 2} \int dx^+ \left\{
(-p_-) (\dot y^2 - \mu^2 y^2 ) - {m^2 \over (-p_-) } \right\}
}}
where in the last line we have chosen lightcone gauge $x^+ = \tau$,
found the conserved quantity $p_- = - e^{-1}$ and plugged it back
into the action. Note that $p_- \leq 0 $. The Hamiltonian
corresponding
to the action \act\ is $H = i \partial_+ = - p_+ $.
We see that we get a harmonic oscillator, possibly with time
dependent frequency. If $\mu$ is constant we get an ordinary
harmonic oscillator. Note that $\mu^2 $ can be negative in
some plane waves. In fact, if the metric \planew\ is a solution of
the vacuum Einstein equations then the trace of $A_{ij}$ is zero, so
that there are both positive and negative eigenvalues. This correspond
to the fact that tidal forces are focusing in one direction and
defocusing in others.

The Klein Gordon equation for a field of mass $m$ can also be
solved by taking  solutions with fixed $p_-$ and then writing the
equation for  constant $\mu$ as
\eqn\kgequation{
 i \partial_+ \phi = \mu (a^\dagger a + 1/2 ) \phi + {m^2 \over 2 (-p_-) }
 \phi
}
For simplicity we consider only one $y$ coordinate.
If $\mu$ is $x^+$ independent then the spectrum is
\eqn\spectr{
-p_+ = \mu ( n+{1\over 2} ) + { m^2 \over 2(- p_-) }
}
The ground state energy is $1/2$ (per dimension)
for a scalar field, but it has
other values of higher spin fields.

People who work in the lightcone gauge sometimes prefer to use
$p^+$ and $p^-$. In our case since we are in a curved space it
is more convenient to stick with $p_\pm$ and live with the inconvenience
that they are typically negative, $p_-$ is always negative.

It turns out that string theory is also solvable on plane waves.
We will discuss only one example.

\subsec{ Type IIB supergravity plane wave}

We will be interested in the following plane wave solution of
IIB supergravity \figueroaiib\
\eqn\pwmetric{
ds^2 = -2 dx^+ dx^-  - y^2 (dx^+)^2  + dy^i dy^i
}
with a constant field strength
\eqn\fs{
F = dx^+ (dy_1 dy_2 dy_3 dy_4  + dy_5 dy_6 dy_7 dy_8)
}
String propagation on this background can be solved exactly.
by choosing light cone gauge in the Green-Schwarz action \refs{\metsaev,\mt}.
The lightcone action becomes
\eqn\lcact{
S = {1 \over 2 \pi \alpha'}
\int dt \int_0^{ \pi  \alpha' |p_-| } d \sigma
\left[ \half \dot y^2 - \half y'^2 - \half \mu^2 y^2
+ i \bar S ( \not \partial  + \mu I) S \right]
}
where $I = \Gamma^{1234}$ and $S$ is a Majorana spinor on the
worldsheet and a positive chirality SO(8) spinor
under rotations in the eight transverse directions.
We quantize this action by expanding all fields in
Fourier modes on the circle labeled by $\sigma$.
For each Fourier mode we get a harmonic oscillator
(bosonic or fermionic depending on the field).
Then the
light cone Hamiltonian is
\eqn\spectrum{
-p_+ = H_{lc} = \sum_{n=-\infty}^{+\infty} N_n \sqrt{
\mu^2 + {  n^2 \over (  \alpha' |p_-|/2)^2}}
}
Here $n$ is the label of the Fourier mode, $n>0$ label
left movers and $n<0$ right movers.  $N_n$ denotes the total
occupation number
of that mode, including bosons and fermions.
Note that the ground state energy of bosonic oscillators is
canceled by that of the fermionic oscillators.
The constraint on the momentum in the sigma direction reads
\eqn\momconstraint{
P = \sum_{n=-\infty}^\infty  n N_n  =0
}

In the limit that $\mu$ is very small, $ \mu \alpha' |p_-|  \ll 1$,
we recover the flat space spectrum.
It is also interesting to consider the opposite limit,
where
\eqn\curvedlim{
 \mu \alpha'  p^+  \gg 1
}
This limit corresponds to strong tidal forces on the strings, i.e.
to strong curvatures.
In this limit all the low lying string oscillator modes have
almost the same energy.
This limit  corresponds to a highly curved
background with RR fields. In fact we will later see that
the appearance of a large number of light modes is
expected from the Yang-Mills theory.
In this limit different pieces of the string move independently.

\ifig\fiftwenty{Strings moving in a plane wave. (a) We see the
weak field limit. (b) correspond to the strong field limit, where
excitations along the string behave as very massive particles.  }
{\epsfxsize 3.5 in\epsfbox{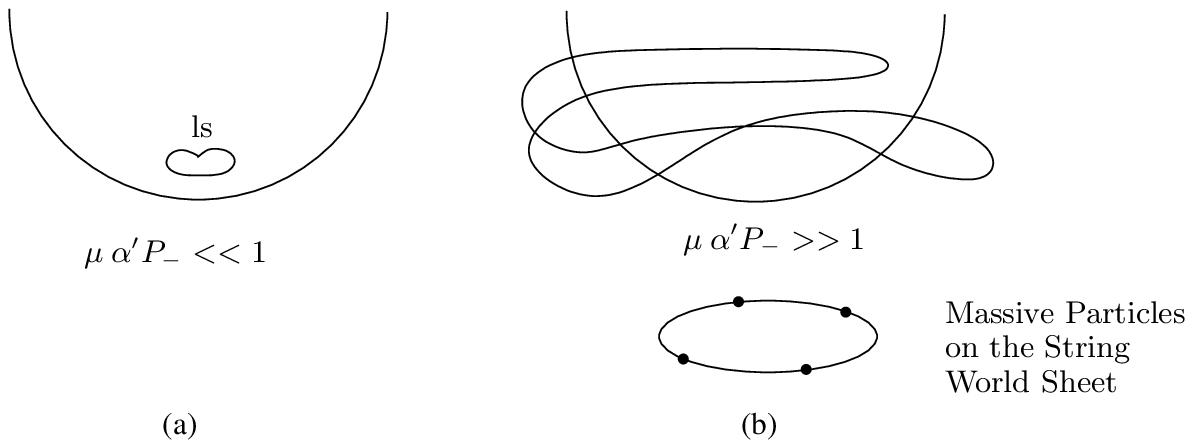}}

\subsec{Type IIB plane wave from $AdS_5 \times S^5$ }

In this subsection we obtain the maximally supersymmetric plane
wave of type IIB string theory
 as a limit of $AdS_5\times S^5$. This is a so called ``Penrose"
 limit. It consists on focusing on the spacetime region near a
 lightlike geodesic.
\ifig\figtwentyone{ In the plane wave limit we focus on the
lightlike trajectory that goes around a great circle of $S^5$ and
sits at the origin of the $AdS$ spatial coordinates.  }
{\epsfxsize 1.5 in\epsfbox{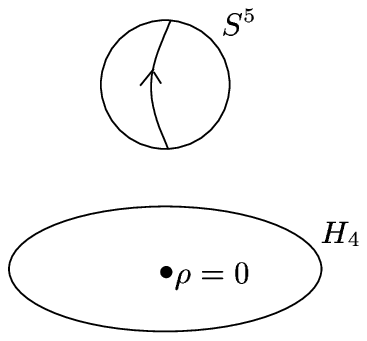}}
The idea is to consider the trajectory of a particle that is moving
very fast along the $S^5$ and to focus on the geometry that this
particle sees. See  \figtwentyone .

We start with the $AdS_5 \times S^5$ metric written as
\eqn\metric{
ds^2 = R^2
\left[ -dt^2 \cosh^2\rho + d\rho^2 + \sinh^2\rho d\Omega_3^2 +
  d\psi^2 \cos^2 \theta  + d\theta^2 + \sin^2 \theta d {\Omega'}^2_3
\right]
}
We want to consider a particle moving along the $\psi$ direction
and sitting at $\rho=0$ and $\theta=0$. We will focus on the
geometry near this trajectory. We can do this  systematically
by introducing coordinates
$\tilde x^- = t - \psi  $, $x^+ = t $ and then
performing the rescaling
\eqn\rescalings{
x^- = R^2 \tilde x^- ~,~~~~~~
\rho = { r \over R} ~,~~~~~~  \theta = { y \over R} ~,~~~~ R \to \infty
}
and $x^+$ is not rescaled. $x^-, ~ r, ~ y $ are kept
fixed as $R \to \infty$.
In this limit the metric \metric\ becomes \ref\BlauDY{
M.~Blau, J.~Figueroa-O'Farrill, C.~Hull and G.~Papadopoulos,
Class.\ Quant.\ Grav.\  {\bf 19}, L87 (2002)
[arXiv:hep-th/0201081].
}
\eqn\metricfin{
ds^2 = - 2 dx^+ dx^- -  ( {\vec r}^{\ 2} + {\vec y}^{\ 2}) (dx^+)^2
+ d {\vec y}^{ \ 2} + d{\vec r}^{\  2}
}
where $\vec y$ and $\vec r$ parametrize points on $R^4$.
We can also see that only the components of $F$ with a plus index
survive the limit.
The mass parameter $\mu$ can be introduced by rescaling
$x^- \to x^-/\mu $ and $x^+ \to  \mu  x^+ $ in eqn. \rescalings .
 These solutions were
studied in \figueroaiib .

It will be convenient for us to understand how the energy and
angular momentum along $\psi$ scale  in the limit \rescalings .
The energy in global coordinates in $AdS$
is given by $E = i \partial_t$ and the
angular momentum by $J = - i \partial_\psi$. This angular
momentum generator can be thought of as the generator
that rotates the 56 plane of  $R^6$.

\subsec{ The ``plane wave" limit in gauge theory variables}

In terms of the dual CFT these are the energy and R-charge
of a state of the field theory on $S^3 \times R$ where
the $S^3$ has unit radius. Alternatively, we can say
that $ E= \Delta$ is the conformal dimension of an
operator on $R^4$.
We find that
\eqn\generators{\eqalign{
 - p_+ =& i \partial_{x^+}
=  i ( \partial_t + \partial_\psi) =  \Delta - J
\cr
 -p_- =& - {\tilde p_- \over R^2} =
 { 1 \over R^2} i \partial_{\tilde x^-}
= { 1 \over R^2} i ( - \partial_\psi) =
{J \over  R^2}
}}

Configurations with fixed non zero
$p_-$
in the limit \rescalings\   correspond to states in $AdS$
 with  large  momentum along the $S^5$, or large $R$ charge
 in the field theory,    $J \sim R^2 \sim (gN)^{1/2}$.
It is useful also to rewrite \spectrum\ in terms of the Yang Mills
parameters. Then we find that the contribution of each
oscillator to $\Delta - J$ is
\eqn\spectrumads{
(\Delta- J)_n =  w_n = \sqrt{1 +  {4 \pi g_s N n^2 \over J^2} }
}
Notice that $g_sN/J^2$ remains fixed in the $g_sN\to \infty $ limit
that we are taking.

When we perform the rescalings
\rescalings\ we can perform the limit in two ways.
If we want to get the plane wave with finite string coupling then
we take the  $N\to \infty$ limit keeping
the string coupling $g_s$ fixed
and we focus on  operators  with $J \sim N^{1/2}$  and
 $\Delta-J$  fixed.

On the other hand  we could first take the 't Hooft limit
$g \to 0$, $gN =$fixed, and then after taking this limit,  we
take the limit of large 't Hooft coupling keeping
$J/\sqrt{gN}$ fixed and $\Delta -J$ fixed.
 Taking the limit in
this fashion gives us a plane wave background with zero string coupling.
Since we will be interested in these notes in the free string spectrum
of the theory it will be more convenient for us to take this second limit.
But to consider string interactions we need to consider the first.

{}From this point of view it is clear that the full supersymmetry
algebra of
the metric \metric\ is a contraction of that of $AdS_5 \times S^5$
\figueroaiib .
This algebra implies that $p_\pm \leq 0$.

\subsec{ Strings from ${\cal N} =4$ Super Yang Mills }

After taking the 't Hooft limit,
we are interested in the limit  of large 't Hooft coupling
$g N \to \infty$.
 We want to consider states
which carry parametrically large
R charge $J \sim \sqrt{gN}$. \foot{Since we first took the
't Hooft limit then  giant gravitons are not
important. }
This R charge generator, $J$,
is the SO(2) generator  rotating two of the six scalar fields.
We want to find the spectrum of states with
$\Delta - J$ finite  in this limit.  We
are interested in single trace states of the Yang Mills theory on
$S^3\times R$, or equivalently, the spectrum of dimensions
of single trace operators of the Euclidean theory on $R^4$.
We will often go back and forth between the states and the
corresponding operators.

Let us first start by understanding the operator  with lowest
value of $\Delta - J=0$. There is a unique single trace
operator with $\Delta -J=0$, namely
$Tr[Z^J]$, where $Z\equiv  \phi^5 + i \phi^6$ and
the trace is over the $N$ color indices.
We are taking $J$ to be the SO(2) generator rotating the plane 56.
At weak coupling the dimension of this operator is $J$ since
each $Z$ field has dimension one.
This operator is a chiral primary and hence its  dimension is protected by
supersymmetry. It is associated to the vacuum state in light cone
gauge, which is the unique state with zero light cone Hamiltonian.
In other words we have the correspondence
\eqn\statezero{
{ 1 \over \sqrt{J}  N^{J/2}} Tr[Z^J]   \Longleftrightarrow
|0,p_+ \rangle_{l.c.}
}
We have normalized the operator as follows.
When we compute $ \langle Tr[\bar Z^J](x)  Tr[Z^J](0) \rangle$
we have $J$ possibilities
for the contraction of the first $\bar Z$ but then planarity
implies that we contract the second $\bar Z$ with a $Z$ that is next
to the first one we contracted and so on. Each of these contraction
gives a factor of $N$. Normalizing this two point function to
one we get the normalization factor in \statezero .\foot{
In general in the free theory any contraction of a single trace
 operator with
its complex conjugate one will give us a factor of $N^{n}$,
where $n$ is the number of fields appearing in the operator. }

Now we can consider other operators that we can build in the free
theory. We can add other fields, or we can add derivatives of
fields like $\partial_{(i_1 } \cdots \partial_{i_n)}\phi^r$, where
we only take the traceless combinations since the traces can be
eliminated via the equations of motion. The order in which these
operators are inserted in the trace is important. All operators
are all ``words'' constructed by these fields up to the cyclic symmetry,
these were discussed and  counted
in  \polyakov .
We will find it convenient to divide all fields, and derivatives of
fields,  that appear
in the free  theory according to their $\Delta - J$ eigenvalue.
There is only one mode that
has $\Delta-J =0$, which is the mode used in \statezero .
There are eight bosonic  and eight fermionic
modes with $\Delta - J =1$. They arise as follows.
First we have the four scalars in the directions not rotated
by $J$, i.e. $\phi^i$, $i=1,2,3,4$. Then we have
derivatives of the field $Z$,  $ D_i Z = \partial_i Z + [A_i,Z]$,
 where $i=1,2,3,4$ are four directions in $R^4$.
Finally  there are  eight fermionic operators $\chi^a_{J=\half}$ which
are the eight components with $J= \half$ of the sixteen
component gaugino $\chi$ (the other eight components have $J=-\half$).
These eight components transform in the positive chirality spinor
representation of $SO(4)\times SO(4)$.
 We will focus first on operators built out
of these fields and then we will discuss what happens when we include
other fields, with $\Delta-J>1$, such as $\bar Z$.

The state \statezero\ describes a particular mode of ten
dimensional  supergravity
in a particular wavefunction \wittenhol .
Let us now discuss how to generate
all other massless supergravity modes.  On the string theory side
we construct all these states by applying the zero momentum oscillators
$a_0^i$, $i=1, \dots ,8$
and $S^b_0$, $b=1, \dots 8$ on the light cone vacuum
 $|0,p_+\rangle_{l.c.}$.
Since the modes on the string are
massive all these zero momentum oscillators
are harmonic oscillators, they all have the same light cone energy.
So the total light cone energy is  equal to the total number of
oscillators that are acting on the light cone ground state.
We know that in  $AdS_5\times S^5 $  all
gravity modes are
in the same supermultiplet
as the state of the form \statezero \ref\GunaydinFK{
M.~Gunaydin and N.~Marcus,
Class.\ Quant.\ Grav.\  {\bf 2}, L11 (1985).
H.~J.~Kim, L.~J.~Romans and P.~van Nieuwenhuizen,
Phys.\ Rev.\ D {\bf 32}, 389 (1985).
}.
The same is clearly true in the
limit that we are considering.
More precisely, the action of all supersymmetries and bosonic
symmetries of the plane wave background (which are intimately
related to the $AdS_5\times S^5$ symmetries) generate all
other ten dimensional massless modes with given $p_-$.
For example, by acting by some of the rotations of $S^5$ that
do not commute with the SO(2) symmetry that we singled out
we create states of the form
\eqn\create{
 {1 \over \sqrt{J} }\sum_{l}  {1 \over \sqrt{J} N^{J/2 +1/2} }
Tr[ Z^l \phi^r Z^{J-l}] =  {1 \over N^{J/2 +1/2} }Tr[\phi^r Z^J]
}
where $\phi^r$, $r=1,2,3,4$  is one of the scalars neutral
under $J$.
In \create\ we used   the cyclicity of the trace.
Note that we have normalized the states appropriately in the
planar limit.
We can act any number of times by these generators and we get
operators  roughly of the form
$\sum Tr[ \cdots Z \phi^r Z\cdots Z \phi^k ] $.
where the sum is over all the possible orderings of the $\phi$s.
We can repeat this discussion with the other $\Delta-J=1$ fields.
Each time we insert a new operator we sum over all possible
locations where we can insert it. Here we are neglecting
possible extra terms that we need when two $\Delta-J=1$ fields
are at the same position, these are subleading in a $1/J$ expansion
and can be neglected in the large $J$ limit that we are considering.
We are also ignoring the fact that $J$ typically decreases when we
act with these operators.
 In other words, when we act with
the symmetries that do not leave $Z$ invariant we will change
one of the  $Z$s in \statezero\ to a field with $\Delta-J=1$,
when we act again with one of the symmetries we can change
one of the $Z$s that was left unchanged in the first step or
we can act on the field that was already changed in the first step.
This second possibility is of lower order in a $1/J$ expansion and
we neglect it. We will always work in a ``dilute gas'' approximation
where most  of the fields in the operator are $Z$s and there
are a few other fields sprinkled in the operator.

For example, a state with two excitations will be of the form
\eqn\statetwo{
\sim { 1 \over N^{J/2 +1}}
  {1\over \sqrt{J} } \sum_{l=0}^J  Tr[\phi^r Z^l \psi_{J=\half}^b
Z^{J-l} ]
}
where we used the cyclicity of the trace to put the $\phi^r$ operator
at the beginning of the expression.
We associate \statetwo\ to the string state
$ a_0^{\dagger k} S^{\dagger \ b}_0 |0,p_+\rangle_{l.c.}$.
Note that for planar diagrams it is very important to keep
track of the position of the operators. For example, two
operators of the form $ Tr[\phi^1 Z^l \phi^2 Z^{J-l} ]$ with
different values of $l$ are orthogonal to each other in the planar limit
(in the free theory).

The conclusion is that there is a precise correspondence between
the supergravity modes and the operators. This is of course
well known \refs{\gkp,\wittenhol,\review}.
Indeed, we see from \spectrum\ that
their $\Delta-J = -p_+ $ is indeed what we compute at weak
coupling, as we expect from the BPS argument.

In order to understand non-supergravity modes in the bulk it is
 clear that what we need to understand  the Yang Mills
description of the states obtained by
the action of the string oscillators which have $n\not = 0$.
Let us consider first one of the string
oscillators which creates a bosonic mode along one of the
four directions that came from the $S^5$, let's say
$a^{\dagger \ 8}_n$. We already understood that the action of
$a^{\dagger \ 8}_0$
corresponds to insertions of an operator $\phi^4$ on all
possible positions along the ``string of $Z$'s''. By a
``string of $Z$s'' we just mean a sequence of $Z$ fields
one next to the other such as we have in \statezero .
We propose that $a_n^{\dagger 8} $ corresponds to the insertion of
the same field $\phi^4$ but now with a position dependent
phase
\eqn\stateop{
  { 1 \over \sqrt{J} } \sum_{l=1}^J  { 1 \over
\sqrt{J} N^{J/2 +1/2}} Tr[ Z^l  \phi^4 Z^{J-l} ]
e^{ 2 \pi i n l \over J}
}
In fact the state \stateop\ vanishes by cyclicity of the trace.
This corresponds to the fact that we have the constraint that
the total momentum along the string should vanish \momconstraint ,
 so that
 we cannot insert only one $a_n^{\dagger \ i}$ oscillator.
So we should
insert more than one  oscillator so that the total momentum
is zero.
For example  we can  consider the string state obtained
by acting with the $a^{\dagger \ 8}_n$ and $a^{\dagger \ 7 }_{-n}$,
which has
zero total momentum along the string. We propose that this state
should be identified with
\eqn\statetwomom{
a^{\dagger \ 8}_n a^{\dagger \ 7 }_{-n} |0,p_+\rangle_{l.c.}
\Longleftrightarrow
 { 1 \over \sqrt{J} } \sum_{l=1}^J  {1 \over N^{J/2 +1}}
 Tr[ \phi^3 Z^l  \phi^4 Z^{J-l} ]
e^{ 2 \pi i n l \over J}
}
where we used the cyclicity of the trace to simplify the expression.
The general rule is pretty clear, for each oscillator mode along
the string we associate one of the $\Delta -J =1$ fields of the
Yang-Mills theory and we sum over the insertions of this field
at all possible positions with a phase proportional to the
momentum.
States whose total momentum is not zero along the string
lead to operators that are
automatically zero by cyclicity of the trace. In this way we
enforce the $L_0 - \bar L_0 =0$ constraint \momconstraint\
 on the string spectrum.

In summary, each string oscillator corresponds to the insertion
of a $\Delta -J =1$ field, summing over all positions with an
$n$ dependent phase, according to the rule
\eqn\summary{\eqalign{
 a^{\dagger i} & \longrightarrow  D_iZ ~~~~{\rm for}~i=1, \cdots, 4
\cr
a^{\dagger j} &  \longrightarrow  \phi^{j-4} ~~~~
{\rm for}~j=5, \cdots, 8
\cr
S^a  &  \longrightarrow  \chi_{J=\half}^a
}}

In order to show that this identification makes sense
we want to compute the conformal dimension, or more
precisely $\Delta - J$,  of these operators at large
't Hooft coupling and show that it matches  \spectrum .
First note that if we set ${gN \over J^2 } \sim 0$ in
\spectrumads\ we find that all modes, independently of $n$
have the same energy, namely one. This is what we find at
weak 't Hooft coupling where all operators of the form
\statetwomom\ have the same energy, independently of $n$.
Expanding the string theory result \spectrumads\  we find that
the first correction is of the form
\eqn\firstcorr{
 (\Delta- J)_n =  w_n = 1 + { 2 \pi g N n^2 \over J^2} + \cdots
}

This looks like a  first order correction in the 't Hooft
coupling and we can wonder if we can reproduce it by a
 a simple perturbative computation.

In order to compute the corrections it is useful to view the
${\cal N} =4$ theory as an ${\cal N}=1$ theory. As an ${\cal N}=1$
theory we have a Yang Mills theory plus three chiral multiplets
in the adjoint representation. We denote these multiplets as
$W^i$, where $i=1,2,3$. We will often set $Z=W^3$ and $W=W^1$.
The theory also has a superpotential
\eqn\superpot{
{\cal W} \sim   g_{YM}  Tr( W^i W^j W^k ) \epsilon_{ijk}
}

The potential for the Yang Mills theory is the sum of two terms,
$V = V_F + V_D$, one coming from $F$ terms and the other from
D-terms. The one coming from $F$ terms arises from the superpotential
and has the form
\eqn\fterms{
V_F \sim \sum_{i j}Tr\left( [W^i,W^j] [\bar W^i,\bar W^j] \right)
}
On the other hand the one coming from $D$ terms has the form
\eqn\dterms{
V_D \sim  \sum_{ij} Tr( [W^i,\bar W^i][W^j,\bar W^j])
}

\ifig\ftermsfig{ Diagrams that come from $F$ terms. The two diagrams
have a relative minus sign. The $F$ terms propagator is a delta function
so that we could replace the three point vertex by a four point
vertex coming from \fterms . If there are no phases in the operator these
contributions vanish. }
{\epsfxsize 2 in\epsfbox{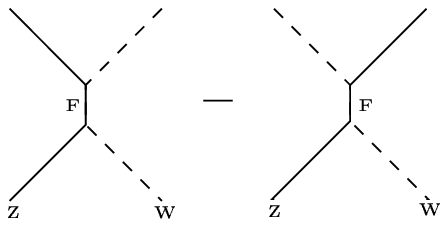}}
\ifig\twentythree{ Diagrams that do not lead to  phase dependence
and cancel out. }
{\epsfxsize 3 in\epsfbox{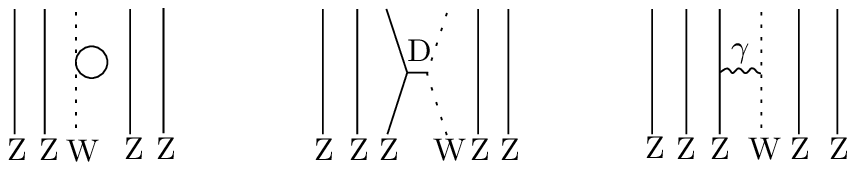}}

We will concentrate in computing the contribution to the
conformal dimension of an operator which contains a $W$ insertion
along the string of $Z$s.
There are various types of diagrams. There are diagrams that
come from $D$ terms, as well as from  photons or self energy corrections.
 There are also
diagrams that come from $F$ terms. The diagrams that come from
$F$ terms can  exchange the $W$ with the $Z$. The $F$ term
contributions cancel in the case that there are no phases, see \ftermsfig.
 This means
that all other diagrams should also cancel, since in the case that there
are no phases we have a BPS object which receives no corrections.
All other one loop diagrams that do not come from $F$ terms do not
exchange the position of $W$, this means that they  vanish also in the
case that there are phases since they will be insensitive to the
presence of phases. In the presence of phases the only diagrams
that will not cancel are then the diagrams that come from the $F$
terms. These are the only diagrams that give a momentum, $n$, dependent
contribution.

In  the free theory, once a $W$ operator is inserted
at one position along the string it will stay there, states with
$W$'s at different positions are orthogonal to each other
in the planar limit (up to the cyclicity of the trace).
We can think of the string of $Z$s in \statezero\ as defining a
lattice, when we insert an operator $W$ at different
positions along the string of $Z$s we are exciting an oscillator
$b^{\dagger}_l$  at the site $l$ on the lattice, $l=1, \cdots J$.
 The interaction
term \fterms\ can  take an excitation from one site
in the lattice to the neighboring site. So we see that the
effects of  \fterms\ will be sensitive to the
momentum $n$.
In fact, one can precisely reproduce \firstcorr\ from \fterms\
including the precise numerical coefficient. Below we give some more
details on the computation.

We will write the square of the
 Yang-Mills coupling
in terms of what in $AdS$ is the string coupling that
transforms as $g \to 1/g$ under S-duality. The trace is just the
usual trace of an   $N\times N$ matrix.

We define $Z = {1 \over \sqrt{2}} ( \phi^5 + i \phi^6)$ and similarly for
$W$.  Then the
propagator is normalized as
\eqn\prop{
\langle Z_{i}^{~j}(x) \bar Z_k^{~l}(0) \rangle =
\delta_i^l \delta_k^j  \ {2 \pi g \over  4 \pi^2} \  { 1 \over  |x|^2}
}
In \fterms\ there is an interaction term  of the form
the form ${ 1 \over {  \pi g}}
\int d^4x   Tr( [Z,W][\bar Z , \bar W])$, where
$W$ is one of the (complex) transverse scalars, let's say $W =W^1$.
The contribution from the $F$ terms shown in \fterms give
\eqn\corr{
<O(x)O^*(0)> = { {\cal N} \over |x|^{2 \Delta} } \left[ 1+ N(4 \pi g)
 (-2+ 2 \cos { 2 \pi n \over J} ) I(x) \right]
}
where ${\cal N}$ is a normalization factor and
 $I(x)$ is the integral
\eqn\integr{
I(x) = {|x|^4 \over  (4 \pi^2)^2} \int d^4 y { 1 \over y^4 (x-y)^4 }
\sim { 1 \over  4 \pi^2} \log|x| \Lambda  + {\rm finite}
}
We extracted the log divergent piece of the integral since it is the
one that
 reflects the change in the  conformal dimension of the operator.

In conclusion we find that for large $J$ and $N$  the first
correction
to the correlator is
\eqn\corr{
<O(x)O^*(0)> = { {\cal N} \over |x|^{2\Delta} } \left[ 1 -
{ 4 \pi g N n^2 \over J^2 } \log(|x|\Lambda) \right]
}
which implies that the contribution of the operator $W$ inserted
in the ``string of $Z$s'' with momentum $n$ gives a contribution to
the anomalous dimension
\eqn\contrib{
(\Delta - J)_n = 1 +  {2 \pi g N n^2 \over J^2}
}
which agrees precisely with the first order term computed from \firstcorr .

There are similar computations we could do for insertions of
$D_iZ$, $\bar W$  or the fermions $\chi^a_{J=1/2}$. In the case of
the fermions the important interaction term will be
a Yukawa coupling of the form $ \bar \chi \Gamma_z [Z \chi] +
\bar \chi \Gamma_{\bar z} [\bar Z , \chi] $. We have not done
these computations explicitly since the 16  supersymmetries
preserved by the state \statezero\
relate them to the computation we did above for the insertion of
 a $W$ operator.

The full square root in \spectrumads\ was derived
in a paper by
Santambrogio and Zanon \ref\zanon{
A.~Santambrogio and D.~Zanon,
arXiv:hep-th/0206079.
}, see also \ref\MinahanVE{
J.~A.~Minahan and K.~Zarembo,
JHEP {\bf 0303}, 013 (2003)
[arXiv:hep-th/0212208].
}.

In summary, the   ``string of $Z$s''  becomes the
physical string and  each $Z$ carries one unit of
$J$ which is one unit of $-p_-$. Locality along the
worldsheet of the string comes from the fact that
planar diagrams  allow  only contractions of neighboring
operators. So the Yang Mills theory gives a string bit
model (see \ref\thorn{
C.~B.~Thorn,
arXiv:hep-th/9405069;
R.~Giles, L.~D.~McLerran and C.~B.~Thorn,
Phys.\ Rev.\ D {\bf 17}, 2058 (1978);
C.~B.~Thorn,
Phys.\ Rev.\ D {\bf 20}, 1934 (1979);
C.~B.~Thorn,
Phys.\ Rev.\ D {\bf 20}, 1435 (1979);
C.~B.~Thorn,
Phys.\ Rev.\ D {\bf 19}, 639 (1979);
R.~Brower, R.~Giles and C.~B.~Thorn,
Phys.\ Rev.\ D {\bf 18}, 484 (1978);
C.~B.~Thorn,
Phys.\ Rev.\ D {\bf 17}, 1073 (1978);
C.~B.~Thorn,
Phys.\ Lett.\ B {\bf 70}, 85 (1977);
R.~Giles and C.~B.~Thorn,
Phys.\ Rev.\ D {\bf 16}, 366 (1977).
})  where each bit is a $Z$ operator. Each bit
carries one unit of $J$ which  is  one unit of $-p_-$.

The reader might, correctly, be thinking  that all
this seems too  good  to be true. In fact, we have  neglected
many other diagrams and many other operators which, at
weak 't Hooft coupling also have small $\Delta-J$.
In particular, we considered operators which arise by
inserting the fields  with $\Delta-J=1$ but we did
not consider the possibility of inserting fields
corresponding to $\Delta - J=2, 3, \dots$, such as
$\bar Z, ~ \partial_k \phi^r, ~ \partial_{(l} \partial_{k)} Z$, etc..
The diagrams of the type we considered above would give rise
to other 1+1 dimensional fields for each of these modes.
These are
present at weak 't Hooft
coupling but they should not be present at
 strong coupling,
since we do not see them  in the string spectrum.
We believe that what happens is that these fields
get a large mass in the $N\to \infty$ limit. In other
words, the operators get a large conformal dimension.
One can compute
the first
correction to the energy (the conformal weight) of the
of the state that results from inserting $\bar Z$ with some
``momentum'' $n$. In contrast to our previous
computation for $\Delta-J=1$ fields we find that
besides an effective kinetic term as in \firstcorr\ there is
an $n$ independent contribution that goes as $gN$ with no
extra powers of $1/J^2$ \bmn .
 This is an indication that these
excitations become very massive in the large $gN$ limit.
In addition, we can compute the decay amplitude of
$\bar Z$ into a pair of $\phi$ insertions. This is also
very large, of order $gN$.

Though we have not done a similar  computation for other fields
with $\Delta -J>1$, we believe that the same will be true
for the other fields.
In general we expect to find many terms in the effective
Lagrangian with  coefficients that are of order
$gN$ with no inverse powers of $J$ to suppress them.
In other words, the  Lagrangian of Yang-Mills on $S^3 $
acting on a state which contains a large number of $Z$s
gives  a Lagrangian on a discretized spatial circle
 with an infinite number
of KK modes. The coefficients of this effective Lagrangian
are factors of $gN$, so all fields will generically get very
large masses.

The only fields that will not get a large mass are those
whose mass is protected for some reason. The fields
with $\Delta-J=1$ correspond to Goldstone bosons and fermions
of the symmetries broken by the state \statezero .
Note that despite the fact that they morally are Goldstone
bosons and fermions, their mass is  non-zero, due to the
fact that the symmetries that are broken do not commute with $p_+$,
the light cone  Hamiltonian. The point is that their masses
are determined, and hence protected, by the (super)symmetry algebra.

Having described how the single string Hilbert space arises it is
natural to ask whether we can incorporate properly the string
interactions.
Clearly string interactions come when we include non-planar diagrams.
There has been a lot of recent work relating the string interactions
to the leading non-planar contributions in Yang Mills
\ref\interactions{
M.~Spradlin and A.~Volovich,
Phys.\ Rev.\ D {\bf 66}, 086004 (2002)
[arXiv:hep-th/0204146].
M.~Spradlin and A.~Volovich,
JHEP {\bf 0301}, 036 (2003)
[arXiv:hep-th/0206073].
D.~Berenstein and H.~Nastase,
arXiv:hep-th/0205048.
N.~R.~Constable, D.~Z.~Freedman, M.~Headrick, S.~Minwalla, L.~Motl, A.~Postnikov and W.~Skiba,
JHEP {\bf 0207}, 017 (2002)
[arXiv:hep-th/0205089].
N.~Beisert, C.~Kristjansen, J.~Plefka, G.~W.~Semenoff and M.~Staudacher,
Nucl.\ Phys.\ B {\bf 650}, 125 (2003)
[arXiv:hep-th/0208178].
C.~Kristjansen, J.~Plefka, G.~W.~Semenoff and M.~Staudacher,
Nucl.\ Phys.\ B {\bf 643}, 3 (2002)
[arXiv:hep-th/0205033].
D.~J.~Gross, A.~Mikhailov and R.~Roiban,
JHEP {\bf 0305}, 025 (2003)
[arXiv:hep-th/0208231].
N.~R.~Constable, D.~Z.~Freedman, M.~Headrick and S.~Minwalla,
JHEP {\bf 0210}, 068 (2002)
[arXiv:hep-th/0209002].
B.~Eynard and C.~Kristjansen,
JHEP {\bf 0210}, 027 (2002)
[arXiv:hep-th/0209244].
H.~Verlinde,
arXiv:hep-th/0206059.
J.~G.~Zhou,
Phys.\ Rev.\ D {\bf 67}, 026010 (2003)
[arXiv:hep-th/0208232].
D.~Vaman and H.~Verlinde,
JHEP {\bf 0311} 044 (2003), arXiv:hep-th/0209215.
Y.~j.~Kiem, Y.~b.~Kim, S.~m.~Lee and J.~m.~Park,
Nucl.\ Phys.\ B {\bf 642}, 389 (2002)
[arXiv:hep-th/0205279].
M.~x.~Huang,
Phys.\ Lett.\ B {\bf 542}, 255 (2002)
[arXiv:hep-th/0205311].
C.~S.~Chu, V.~V.~Khoze and G.~Travaglini,
JHEP {\bf 0206}, 011 (2002)
[arXiv:hep-th/0206005].
P.~Lee, S.~Moriyama and J.~w.~Park,
Phys.\ Rev.\ D {\bf 66}, 085021 (2002)
[arXiv:hep-th/0206065].
M.~x.~Huang,
Phys.\ Rev.\ D {\bf 66}, 105002 (2002)
[arXiv:hep-th/0206248].
P.~Lee, S.~Moriyama and J.~w.~Park,
Phys.\ Rev.\ D {\bf 67}, 086001 (2003)
[arXiv:hep-th/0209011].
R.~de Mello Koch, A.~Jevicki and J.~P.~Rodrigues,
arXiv:hep-th/0209155.
R.~A.~Janik,
Phys.\ Lett.\ B {\bf 549}, 237 (2002)
[arXiv:hep-th/0209263].
J.~H.~Schwarz,
JHEP {\bf 0209}, 058 (2002)
[arXiv:hep-th/0208179].
A.~Pankiewicz,
JHEP {\bf 0209}, 056 (2002)
[arXiv:hep-th/0208209].
I.~R.~Klebanov, M.~Spradlin and A.~Volovich,
Phys.\ Lett.\ B {\bf 548}, 111 (2002)
[arXiv:hep-th/0206221].
J.~Pearson, M.~Spradlin, D.~Vaman, H.~Verlinde and A.~Volovich,
JHEP {\bf 0305}, 022 (2003)
[arXiv:hep-th/0210102].
J.~Gomis, S.~Moriyama and J.~w.~Park,
Nucl. Phys. B { \bf 659} 179 (2003), arXiv:hep-th/0210153.
J.~Gomis, S.~Moriyama and J.~w.~Park,
Nucl.\ Phys.\ B {\bf 665}, 49 (2003)
[arXiv:hep-th/0301250].
J.~Gomis, S.~Moriyama and J.~w.~Park,
arXiv:hep-th/0305264.
A.~Pankiewicz and B.~.~Stefanski,
Nucl.\ Phys.\ B {\bf 657}, 79 (2003)
[arXiv:hep-th/0210246].
B.~.~Stefanski,
arXiv:hep-th/0304114.
N.~Mann and J.~Polchinski,
arXiv:hep-th/0305230.
K.~Skenderis and M.~Taylor,
JHEP {\bf 0206}, 025 (2002)
[arXiv:hep-th/0204054].
C.~S.~Chu and V.~V.~Khoze,
JHEP {\bf 0304}, 014 (2003)
[arXiv:hep-th/0301036].
N.~Beisert, C.~Kristjansen, J.~Plefka and M.~Staudacher,
Phys.\ Lett.\ B {\bf 558}, 229 (2003)
[arXiv:hep-th/0212269].
Y.~j.~Kiem, Y.~b.~Kim, J.~Park and C.~Ryou,
JHEP {\bf 0301}, 026 (2003)
[arXiv:hep-th/0211217].
P.~Di Vecchia, J.~L.~Petersen, M.~Petrini, R.~Russo and A.~Tanzini,
arXiv:hep-th/0304025.
A.~Pankiewicz,
JHEP {\bf 0306}, 047 (2003)
[arXiv:hep-th/0304232].
D.~Z.~Freedman and U.~Gursoy,
JHEP {\bf 0308}, 027 (2003)
[arXiv:hep-th/0305016].
G.~Georgiou and V.~V.~Khoze,
JHEP {\bf 0304}, 015 (2003)
[arXiv:hep-th/0302064].
U.~Gursoy,
 JHEP {\bf 0310} 027 (2003), arXiv:hep-th/0212118.
C.~S.~Chu, M.~Petrini, R.~Russo and A.~Tanzini,
Class.\ Quant.\ Grav.\  {\bf 20}, S457 (2003)
[arXiv:hep-th/0211188].
Y.~H.~He, J.~H.~Schwarz, M.~Spradlin and A.~Volovich,
Phys.\ Rev.\ D {\bf 67}, 086005 (2003)
[arXiv:hep-th/0211198].
R.~Roiban, M.~Spradlin and A.~Volovich,
 JHEP {\bf 0310} 055 (2003), arXiv:hep-th/0211220.
C.~S.~Chu, V.~V.~Khoze, M.~Petrini, R.~Russo and A.~Tanzini,
arXiv:hep-th/0208148.
}.

Finally we should note that there is another interesting limit
where we consider operators with large spin
\ref\GubserTV{
S.~S.~Gubser, I.~R.~Klebanov and A.~M.~Polyakov,
Nucl.\ Phys.\ B {\bf 636}, 99 (2002)
[arXiv:hep-th/0204051].
}. In this case one finds that for large spin the operators
have dimensions $\Delta - S \sim (const) \sqrt{ g^2 N} \log S$.
At weak coupling one has a similar relation but
 in front of the logarithm we have a factor of $g^2 N$.

\bigskip\bigskip\bigskip
\noindent {\bf Acknowledgments:}

It is a pleasure to thank the local organizers and lecturers of
the TASI 03 school. I would also like to thank  A. Kobrinskii,
G. Moore and A. Vainshtein for
pointing out several typos in a previous version.

This work was supported in part by DOE grant DE-FG02-90ER40542.

\listrefs

\bye